\documentclass[12pt,preprint]{aastex}

\usepackage{subfigure}

\newcommand{\rCore}{R_{\mbox{\tiny{core}}}}

\newcommand{\rhoCore}{\rho_{\mbox{\tiny{core}}}}
\newcommand{\RT}{R_{\mbox{\tiny{T}}}}

\newcommand{\etal}{et~al.}
\newcommand{\Msun}{\mbox{${\rm M}_{\odot}$}}

\newcommand{\BH}{\mbox{\tiny{BH}}}
\newcommand{\jetADAF}{\mbox{\tiny{jet,ADAF}}}
\newcommand{\jetTD}{\mbox{\tiny{jet,TD}}}
\newcommand{\cool}{\mbox{\tiny{cool}}}
\newcommand{\vir}{\mbox{\tiny{vir}}}
\newcommand{\disk}{\mbox{\tiny{disk}}}
\newcommand{\virGas}{\mbox{\tiny{vir,gas}}}
\newcommand{\dynHalo}{\mbox{\tiny{dyn,halo}}}
\newcommand{\dynDisk}{\mbox{\tiny{dyn,disk}}}
\newcommand{\hot}{\mbox{\tiny{hot}}}
\newcommand{\cold}{\mbox{\tiny{cold}}}
\newcommand{\crit}{\mbox{\tiny{crit}}}
\newcommand{\halo}{\mbox{\tiny{halo}}}
\newcommand{\SN}{\mbox{\tiny{SN}}}
\newcommand{\ejected}{\mbox{\tiny{ejected}}}
\newcommand{\returned}{\mbox{\tiny{returned}}}
\newcommand{\heated}{\mbox{\tiny{heated}}}
\newcommand{\cocoon}{\mbox{\tiny{cocoon}}}
\newcommand{\shock}{\mbox{\tiny{shock}}}
\newcommand{\acc}{\mbox{\tiny{acc}}}
\newcommand{\rad}{\mbox{\tiny{rad}}}
\newcommand{\kms}{\mbox{km\,s}$^{-1}$}
\newcommand{\kmsInside}{\mbox{km\,s}^{-1}}
\newcommand{\kpcInside}{\mbox{kpc}}
\newcommand{\DM}{\mbox{\tiny{DM}}}
\newcommand{\gas}{\mbox{\tiny{gas}}}
\newcommand{\dyn}{\mbox{\tiny{dyn}}}

\newcommand{\rec}{\mbox{\tiny{rec}}}
\newcommand{\reheated}{\mbox{\tiny{reheated}}}
\newcommand{\shocked}{\mbox{\tiny{shocked}}}

\newcommand{\edd}{\mbox{\tiny{edd}}}

\newcommand{\jetMin}{\mbox{\tiny{jet,min}}}
\newcommand{\core}{\mbox{\tiny{core}}}

\newcommand{\critDown}{\mbox{\tiny{crit,down}}}
\newcommand{\critUp}{\mbox{\tiny{crit,up}}}
\newcommand{\bulge}{\mbox{\tiny{bulge}}}

\shorttitle{Radio source feedback in galaxy evolution}
\shortauthors{Shabala \& Alexander}

\begin{document}

\title{Radio source feedback in galaxy evolution}


\author{Stanislav Shabala$^{1,2}$ and Paul Alexander$^{1,3}$}
\affil{$^1$ Cavendish Astrophysics, University of Cambridge, JJ Thomson Avenue, Cambridge CB3 0HE\\
$^2$ Oxford Astrophysics, Denys Wilkinson Building, Keble Road, Oxford OX1 3RH\\
$^3$ Kavli Institute for Cosmology, Madingley Road, Cambridge CB3 0HA}
\email{stas.shabala@astro.ox.ac.uk.}

\begin{abstract}

We present a galaxy evolution model which incorporates a physically motivated implementation of AGN feedback. Intermittent jets inflate cocoons of radio plasma which then expand supersonically, shock heating the ambient gas. The model reproduces observed star formation histories to the highest redshifts for which reliable data exists, as well as the observed galaxy colour bimodality. Intermittent radio source feedback also naturally provides a way of keeping the black hole and spheroid growth in step. We find possible evidence for a top-heavy Initial Mass Function (IMF) for $z>2$, consistent with observations of element abundances, and sub-mm and Lyman break galaxy counts.

\end{abstract}


\section{Introduction}

A remarkably simple and self-consistent picture of structure formation via growth of the gravitational instability has emerged over the past few decades.  Recent measurements of the cosmic microwave background (CMB) power spectrum by the Wilkinson Microwave Anisotropy Probe (WMAP; Spergel \etal\/ 2007), and the galaxy clustering power spectrum observations from the two degree Field Galaxy Redshift Survey (2dFGRS; Colless \etal\/ 2001) and the Sloan Digital Sky Survey (SDSS; York \etal\/ 2000) have confirmed the cold dark matter model with a non-zero cosmological constant (the $\Lambda$CDM model) as providing the current state-of-the art description of structure formation. This cosmogony is consistent with the rate of expansion of the Universe as inferred from supernovae observations, present mass budget of the Universe, observed baryonic fraction in rich clusters, structure of the Lyman-$\alpha$ forest at $z=3$, and the model of Big Bang nucleosynthesis (Spergel \etal\/ 2007).

Structures grow via amplification of overdensities. Gunn \& Gott (1972) were the first to use the spherical top-hat model to track the growth of clusters. Press \& Schechter (1974) extended this analysis to growth of Gaussian fluctuations by smoothing the field on different scales to predict the dark matter halo mass function. The results turned out to be in remarkable agreement with later numerical simulations that followed the evolution of structure subject to the gravitational instability (e.g. Lacey \& Cole 1994; Somerville \etal\/ 2000). Sheth \etal\/ (2001) refined the Press-Schechter formalism (to the so-called Extended Press-Schechter formalism, EPS) to consider ellipsoidal, rather than spherical, collapse. The merger histories of the dark matter haloes can be extracted by employing either percolation algorithms or Monte-Carlo simulations. With the merger trees in place, it makes sense to talk about individual halo accretion histories. Analytical fits to halo mass functions (e.g. Jenkins \etal\/ 2001) and average mass accretion histories of individual haloes of a given mass at the present epoch (e.g. van den Bosch 2002) are now available. In addition, the density and temperature structure of dark matter haloes has been studied extensively using numerical simulations.

Observable galaxy properties are due to the baryonic component of these structures. White \& Rees (1978) suggested that haloes accrete baryonic matter along with the dark matter. The baryons are shock heated to the virial temperature of the halo, $T_{\rm vir}$, and thermalization of dark matter particles supports the halo from further collapse. As the hot gas cools, it is no longer supported by radiative pressure, and sinks to the centre of the halo potential. Once cold, the gas suffers from the gravitational instability, and collapses to form stars (Kennicutt 1989).

While very attractive, this picture is at odds with the observable galaxy properties. In the hierarchical assembly paradigm lower mass galaxies form first, and it is the most massive galaxies that should be undergoing vigorous star formation at the present epoch. However, observations (e.g. Silk 2002 and references therein) show these massive galaxies to be systematically redder (as calculated from their optical colours), implying stars in these systems are older, than their less massive counterparts. Infra-red colours of these objects are consistent with passive evolution since $z \sim 1$ \citep{BenderSaglia99}. Using the rest-frame UV flux as a proxy for instantaneous star formation rate, derived star formation rates peak at $z \sim 1-2$ \citep{GiavaliscoEA04}, consistent with this picture. Thus, in this ``cosmic downsizing'' scenario \citep{CowieEA96} the higher mass galaxies are assembled first.

Further evidence for suppression of star formation in massive galaxies comes from observations of the local optical luminosity function. Under the assumption of a constant mass-to-light ratio, the number counts of the most massive and least massive galaxies are significantly overpredicted when compared with observations (Baugh~2006; Eke \etal\/ 2006). While reionization and supernovae feedback provide sufficient heating to reconcile counts at the faint end of the luminosity function, this is insufficient in the most massive galaxies.

The overprediction of stellar content in the most massive structures at the present epoch is closely related to another well-known problem. The cores of galaxy clusters are dense, and hence the gas there has very short cooling times compared to the Hubble time. This originally led to the idea of a cooling flow, and the expectation of large amounts of cold gas being deposited in such cores. However, recent {\em{Chandra}} and {\em{XMM-Newton}} spectroscopic observations (Tamura \etal\/ 2001; Peterson \etal\/ 2003) have found no evidence of X-ray gas cooling below $\sim T_{\rm vir} /3$, suggesting a temperature floor of about 1~keV. Numerous scenarios have been envisaged in attempts to explain these observations. It has been argued that the gas may be cooling, but the line emission is either absorbed, or greatly diminished through intrinsic X-ray absorption \citep{PetersonEA01}, or due to the presence of multiphase gas and inhomogeneity of iron abundance \citep{FabianEA01}. An alternative to hiding the signatures of cooling gas is some form of non-gravitational heating.

In order to offset the cooling in cluster cores, such heating must come from a central source. Burns (1990) found that 70 percent of cD galaxies contain radio sources at their centres. Observations of strong interaction between these objects and the X-ray gas \citep{FabianEA03,FormanEA05,McNamaraEA05}, as well as numerical simulations \citep{ChurazovEA01,BassonAlexander03} suggest feedback from radio-loud Active Galactic Nuclei (AGNs) can have a profound impact on their surroundings. Moreover, a number of tight correlations exist between the global galaxy properties such as spheroid mass \citep{MagorrianEA98,HaeringRix04} and velocity dispersion \citep{GebhardtEA00}, and those of the black holes found at their centres.

AGN feedback provides a natural link between the black hole properties and those of the host galaxy, and has been invoked by a number of researchers to explain the disagreements between the hierarchical structure assembly paradigm and observations. Granato \etal\/ (2004), Bower \etal\/ (2006), Cattaneo \etal\/ (2006) and Croton \etal\/ (2006) have all investigated the role of feedback from radio galaxies in this process. These authors found that the feedback can suppress gas cooling and hence star formation significantly in massive hosts. However, the feedback models they used were largely phenomenological. Croton \etal\/ assumed a rather arbitrary formulation for the rate of black hole fuelling, thus ensuring star formation is preferentially suppressed in massive galaxies. Cattaneo \etal\/ similarly assumed that AGNs only switch on in haloes more massive than some critical value. Granato \etal\/ assumed a constant fraction of AGN kinetic luminosity couples to the gas, and derived approximate kinetic luminosities from known scalings of observable quantities. Bower \etal\/ postulated that cooling in galaxies is suppressed when the radiative output of the cooling gas exceeds some fraction (typically $\sim 0.5$) of the Eddington luminosity.

While encouraging, these approaches do not provide a realistic representation of the interaction between the radio jets and the surrounding gas. The aforementioned authors have to a large degree built into their models factors that guarantee the success of the AGN feedback process. Thus, whether or not radio galaxies really can provide sufficient heating to quench gas cooling by the required amount remains an open question. We address this issue by developing and applying a galaxy evolution model that, for the first time, incorporates detailed and physically motivated, radio source feedback.

AGN feedback is an inherently intermittent process. Heating of the intracluster gas (ICM) decreases the rate of accretion onto the central black hole, until it is shut off completely (e.g. Shabala \etal\/ 2008). Once the gas has had sufficient time to cool, the accretion can restart. It has been suggested (Jester 2005 and references therein) that accretion processes in Galactic X-ray black hole binaries (BHXBs) are analogous to those in AGNs, albeit on much shorter timescales. At least two different accretion states are observed in BHXBs; these are characterised by different spectral shapes and luminosities. The high luminosity/soft X-ray (thermal) spectrum state is identified with the standard Shakura \& Sunyaev (1973) thin accretion disk. The resultant accretion flow is optically thick and radiatively efficient. By constrast, the low luminosity/hard X-ray spectrum state is identified with an optically thin accretion flow. The cooling times are long and trapped radiation is advected inward, resulting in a ``puffed'' disk \citep{ParkOstriker01}. The resulting quasi-spherical Advection Dominated Accretion Flow (ADAF; Narayan \& Yi 1995) is thus radiatively inefficient, and the BHXB is observed in a low luminosity state. Radio jets are only observed in the ADAF phase, with quenching occurring when the BHXB enters the high/soft (i.e. thin disk) state \citep{GalloEA03}.

In this paper, we develop a galaxy evolution model that includes intermittent radio source feedback. In our model, existence of radio jets depends on the nature of the accretion flow solution, which in turn is sensitive to the interplay between cooling and heating of the cluster gas. We describe the major features of our galaxy formation model in Section~\ref{sec:coolingModel}, and incorporate feedback in Section~\ref{sec:feedbackProcesses}. Constraints on the model are placed in Section~\ref{sec:parameterSpace}, and it is tested by means of a comparison with the statistical properties of the observed galaxy population in Section~\ref{sec:results}.

Throughout the paper, we adopt a flat cosmology of $\Omega_{\rm M}=0.25$, $\Omega_\Lambda=0.75$, $h=0.7$ and $\sigma_8 = 0.9$, consistent with the 2dFGRS \citep{CollessEA01} and WMAP \citep{SeljakEA05} results, and the Millennium Simulation \citep{SpringelEA05}.

\section{Basic framework}
\label{sec:coolingModel}

\subsection{Dark matter halo evolution}
\label{sec:haloEvolution}

To follow the evolution of an individual dark matter halo, the analytical description of van den Bosch (2002) is employed. Van den Bosch uses an extended Press-Schechter formalism to derive halo mass accretion histories which are in good agreement with the $\Lambda$CDM simulations of Kauffmann \etal\/ (1999a). The average mass accretion history of a halo with mass $M$ in the local Universe is given by

\begin{equation}
  \log \left< \frac{M(z)}{M} \right> = -0.301 \left[ \frac{\log (1+z)}{\log (1+z_{\rm f})}^{\nu_{\rm f}} \right] ,
\label{eqn:avgMAH}
\end{equation}
where the fitting parameters $\nu_{\rm f}$ and $z_{\rm f}$ are a function of halo mass and cosmology.

For the adopted flat cosmology, $z_{\rm f}$ is obtained by solving

\begin{equation}
  \frac{\left[ \Omega (z_{\rm f}) \right]^{0.0055}}{D (z_{\rm f})} = \Omega_M^{0.0055} + 0.283 \left[ 2(\sigma^2(0.254 M)-\sigma^2(M)) \right]^{0.5} .
\label{eqn:Omegaz}
\end{equation}

Here, $\Omega (z) = \frac{\Omega_{\rm M} (1+z)^3}{\Omega_{\Lambda} + \Omega_{\rm M} (1+z)^3}$; and $D(z)$ is the linear growth factor normalised to unity at $z=0$. Using the fit of Eisenstein \& Hu (1999),

\begin{equation}
  D(z) = \frac{\Omega(z) \left[ \Omega_{\rm M}^{4 \over 7} - \Omega_\Lambda + (1+\frac{\Omega_{\rm M}}{2})(1+\frac{\Omega_\Lambda}{70}) \right]}{(1+z) \left[ \Omega(z)^{4 \over 7} - \left( \frac{\Omega_\Lambda}{\Omega_{\rm M}} \right) \frac{\Omega(z)}{1+z} + (1+\frac{\Omega(z)}{2})(1+\left( \frac{\Omega_\Lambda}{\Omega_{\rm M}} \right) \frac{\Omega(z)}{70(1+z)}) \right]}
\label{eqn:linearGrowthFactor} .
\end{equation}

Following van~den~Bosch (2002), another fitting function is adopted for $\sigma(M)$,

\begin{equation}
  \sigma(M) = \sigma_8 \frac{f \left(3.084 \times 10^{-4} \Gamma \left[ \frac{M}{\Omega_{\rm M}} \right]^{1 \over 3} \right)}{f(32 \Gamma)} ,
\label{eqn:sigmaM}
\end{equation}
where $\Gamma=0.15$ is the power spectrum shape parameter. The function $f$ is

\begin{equation}
  f(u)=64.087 \times (1+1.074 u^{0.3}-1.581 u^{0.4}+0.954 u^{0.5}-0.185 u^{0.6})^{-10} .
\label{eqn:fu}
\end{equation}

Equations \ref{eqn:Omegaz}-\ref{eqn:fu} allow $z_{\rm f}$ to be determined, given a halo mass at redshift zero and cosmology. The second parameter describing the average halo mass accretion history is then given by

\begin{equation}
  \nu_{\rm f} = 1.211+1.858 \log (1+z_{\rm f})+0.308 \Omega_\Lambda^2-0.032 \log \left( \frac{M}{10^{11} h^{-1} M_\odot} \right) .
\label{eqn:nuf}
\end{equation}

Together, equations \ref{eqn:avgMAH}-\ref{eqn:nuf} uniquely determine the average mass accretion history of a halo with a given mass at redshift zero.

Detailed N-body simulations for the $\Lambda$CDM cosmology (e.g. Jenkins \etal\/ 1998; Kauffmann \etal\/ 1999a) provide the present-day halo mass function. Analytical fits to the results of the Millennium Simulation \citep{SpringelEA05} provide a halo mass function at each step of the evolution. Since our model does not explicitly follow mergers, such halo mass functions are required in order to predict galaxy population properties as a function of cosmic epoch.

\subsection{Galaxy formation and evolution}
\label{sec:baryonEvolution}

\subsubsection{Density profiles}
\label{sec:gasDensityProfile}

Gas will be accreted onto the halo along with the dark matter (e.g. White \& Frenk 1991; Scannapieco \etal\/ 2005). We take the mass of baryonic matter accreted at each point in the halo's evolution to be a fraction $f_{\rm b}$ of the total accreted mass. For high mass haloes $f_{\rm b} \approx \frac{\Omega_{\rm B}}{\Omega_{\rm M}}$. Photoionization heating from the UV background reduces this fraction in low-mass haloes; this is discussed in Section~\ref{sec:photoionization}. The gas is shocked to its virial temperature and density as it is accreted at the halo virial radius,

\begin{eqnarray}
  \rho_{\virGas} & = & \Delta_{\vir} \frac{3 H_{\rm 0}^2}{8 \pi G} \Omega_{\rm B} (1+z)^3 \nonumber \\
  R_{\vir} & = & \left[ \frac{3 M_{\vir}}{4 \pi \rho_{\virGas} \left( \frac{\Omega_{\rm M}}{\Omega_{\rm B}} \right)} \right]^{1/3} \\
  T_{\vir} & = & \frac{1}{2} \frac{G M_{\vir}}{R_{\vir}} \frac{\mu m_{\rm H}}{k_{\rm b}} , \nonumber
\label{eqn:rhoTvir}
\end{eqnarray}
where $\Delta_{\vir} = 18 \pi^2 + 82 (\Omega (z)-1)-39(\Omega (z)-1)^2$ \citep{BryanNorman98} is the mean overdensity at the virial radius; and $\mu m_{\rm H}$ is the mean particle mass, with $\mu=0.62$ for a fully ionised gas. The halo virial velocity is
\begin{equation}
  V_{\vir} = \left( \frac{G M_{\vir}}{R_{\vir}} \right)^{1/2} .
  \label{eqn:Vvir}
\end{equation}

The potential well of a spherical dark matter halo is assumed to follow the Navarro, Frenk and White (NFW; Navarro, Frenk \& White 1997) profile,
\begin{equation}
  \rho_{\DM}(r)=\frac{\rho_{{\DM},0}}{\left( \frac{r}{R_{\rm s}} \right) \left( 1+\frac{r}{R_{\rm s}} \right)^2} ,
\label{eqn:rhoDM}
\end{equation}
where scale factor $R_{\rm s}=\frac{R_{\vir}}{C}$ for concentration index $C=\frac{8}{1+z} \left( \frac{M_{\vir}}{1.4 \times 10^{14} M_{\odot}} \right)^{-0.13}$ \citep{BullockEA01}. Central density $\rho_{{\DM},0}$ is set by the constraint that halo mass $M_{\halo}$ is contained within the virial radius. The NFW profile is amenable to integration in closed form, yielding
\begin{equation}
  \rho_{{\DM},0}=\left( \frac{\Omega_{\rm M}-\Omega_{\rm B}}{\Omega_{\rm M}} \right) \left( \frac{M_{\halo}}{4 \pi R_{\vir}^3 f \left( \frac{1}{C} \right)} \right) ,
\label{eqn:rho0}
\end{equation}
where $f(x)=x^3 \left[ {\rm ln} \left( 1+\frac{1}{x} \right) - \frac{1}{1+x} \right]$.

We assume the density distribution of hot X-ray gas also follows Equation~\ref{eqn:rho0}. As with the dark matter, the constraint that all the gas is located within the virial radius determines the central density $\rho_{{\gas},0}$. In this fashion, the density profile can be determined at each point in the halo's evolution.

\subsubsection{Gas cooling and star formation}
\label{sec:coolingAndSF}

The gas subsequently cools, with the cooling time of each gas parcel being given by $t_{\cool} = \frac{3}{2} \frac{\rho_{\gas} k_{\rm B} T}{\mu m_{\rm H} n_e^2 \Lambda(T,Z)}$, where electron density $n_e = \frac{2+\mu}{5 \mu} \frac{\rho_{\gas}}{m_{\rm H}}$ for a fully ionised gas. This yields

\begin{equation}
  t_{\cool} = \frac{75 \mu}{2 (2+\mu)^2} \frac{m_{\rm H} k_{\rm B} T}{\rho_{\gas} \Lambda(T,Z)} .
\label{eqn:tCool}
\end{equation}
The cooling function $\Lambda(T,Z)$ depends on gas temperature and metallicity. Detailed cooling models of Sutherland \& Dopita (1993) for $Z=0.1 Z_{\odot}$ are used here.

The cool gas is deposited onto an accretion disk after a dynamical time $t_{\dyn} (r) = \frac{r}{V_{\vir}}$, and subsequently star formation can take place. Kennicutt (1989) showed that star formation rates in a sample of nearby spiral galaxies have a sharp cutoff below a certain surface density, in line with gravitational instability considerations. This critical surface density is representative of the disk gas density (except at the very edges), and thus one can define a critical disk mass below which no star formation takes place. Kauffmann (1996) gives the threshold density as $\Sigma_{\crit} = 0.59 \left( \frac{V_{\vir}}{\kmsInside} \right) \left( \frac{R_{\disk}}{\kpcInside} \right)$~M$_{\odot}$\,pc$^{-2}$. Integrating over the surface of the disk, the critical mass is $M_{\crit} = 7.5 \times 10^8 \left( \frac{V_{\vir}}{\kmsInside} \right) \left( \frac{R_{\disk}}{\kpcInside} \right)$~M$_{\odot}$. Adopting disk radius $R_{\disk}=0.1 R_{\vir}$ \citep{Kauffmann96,CrotonEA06}, this yields

\begin{equation}
  M_{\crit}=7.5 \times 10^7 \left( \frac{V_{\vir}}{\kmsInside} \right) \left( \frac{R_{\vir}}{\kpcInside} \right)~M_{\odot} .
  \label{eqn:MdiskThresh}
\end{equation}

Following Croton \etal\/ (2006), the instantaneous star formation rate is taken to be
\begin{equation}
  \dot{M}_{\star}= \alpha_{\rm SF} \frac{M_{\cold}-M_{\crit}}{t_{\dynDisk}} ,
  \label{eqn:SFR}
\end{equation}
where disk dynamical time is $t_{\dynDisk}=\frac{R_{\disk}}{V_{\vir}}$ and star formation efficiency $\alpha_{\rm SF} \approx 0.1$ (Croton \etal\/ 2006). Here, stellar mass refers to the total stellar mass within the dark matter halo.

\subsection{Black hole growth}
\label{sec:BHgrowth}

In standard quasar models the build up of black hole mass occurs via both accretion and mergers. In the present model the {\it average} accretion history of a given halo is considered, and thus mergers are implicitly included. When two progenitors merge, we assume the black holes also merge to form a new object with mass equal to the sum of the progenitor masses (e.g. Kauffmann \& Haehnelt 2000; Croton \etal\/ 2006). Thus the contribution from mergers to black hole mass build up follows the growth of the dark matter halo.

Black holes will also grow through accretion of cold disk gas. Clearly, the growth rate will strongly depend on the availability of the cold gas. The shallower gravitational potential wells of lower-mass haloes make gas there more susceptible to ejection by supernovae and AGNs. Following Kauffmann \& Haehnelt (2000; see also Malbon \etal\/ 2007 and Lagos \etal\/ 2008) and parametrising the accreted gas fraction by halo mass, the mass accreted onto the black hole in a time $\small \Delta t$ is
\begin{equation}
	\Delta M_{\BH} = \frac{\epsilon_{\acc} M_{\cold}}{1+(280\,{\rm km\,s^{-1}} / V_{\vir})^2} .
	\label{eqn:MdotBH}
\end{equation}

\section{Feedback processes}
\label{sec:feedbackProcesses}

\subsection{Reionization}
\label{sec:photoionization}

At low halo masses, gas accretion is suppressed due to the ionizing radiation from extragalactic UV background photons. Gnedin (2000) modelled this effect by introducing a filtering mass $M_{\rm f}$ below which the accreted gas fraction $f_{\rm b}$ is reduced from the universal value $f_{\rm b,0}=\frac{\Omega_{\rm B}}{\Omega_{\rm M}}$,

\begin{equation}
  f_{\rm b}(z,M_{\vir}) = \frac{f_{\rm b,0}}{\left[ 1+0.26 \left( \frac{M_{\rm {f}}(z)}{M_{\vir}} \right) \right]^3} .
  \label{eqn:fracBaryon}
\end{equation}

Semi-analytic models typically assume accretion is suppressed when some measure of gas temperature (e.g. the average temperature, temperature of dense clumps or at the periphery of the halo) exceeds the virial temperature of the halo. Okamoto \etal\/ (2008a) showed that such an approach overstates the effects of photoionization feedback, and it is necessary to follow the full merger history of the haloes. Since our model does not explicitly follow mergers, we adopt the Okamoto \etal\/ (2008a) results for the filtering mass as a function of redshift. As is discussed in Section~\ref{sec:parameterSpace}, the effects of this photoionization heating are negligible at halo masses above $10^{12}$~$M_{\odot}$.

\subsection{Supernovae feedback}
\label{sec:SNfeedback}

Newly formed massive stars end their lives as supernovae. The supernova outbursts inject a significant amount of energy into the surrounding gas, reheating the cold disk material and, in the case of the most powerful events, uplifting the gas out of the cluster. They also inject heavy elements into the gas. We follow the treatment of De~Lucia \etal\/ (2004) and Croton \etal\/ (2006) in modelling these processes.

At each point in time, a fraction of the stellar mass formed is instantaneously recycled into cold disk gas. The exact fraction depends on the assumed Initial Mass Function (IMF). Consistent with the adopted diet Salpeter IMF (see Section~\ref{sec:derivedQuantities}) we take $f_{\rec}=0.33$.

The amount of reheated disk gas is proportional to the recycled mass, and hence to the mass of stars formed in the timestep of interest,
\begin{equation}
  \Delta M_{\reheated} = \epsilon_{\disk} \Delta M_{\star} ,
  \label{eqn:dMreheated}
\end{equation}
where $\epsilon_{\disk} = 3.5$ \citep{CrotonEA06}. This raises the total thermal energy of the hot gas by
\begin{equation}
  \Delta E_{\hot} = \frac{1}{2} \Delta M_{\reheated} V_{\vir}^2 .
  \label{eqn:dEhot}
\end{equation}
The energy output of a supernova event is
\begin{equation}
  \Delta E_{\SN} = \frac{1}{2} \epsilon_{\halo} \Delta M_{\star} V_{\SN}^2 ,
  \label{eqn:dEsn}
\end{equation}
where $V_{\SN}=960$~\kms\/ is found by equating $\frac{1}{2} V_{\SN}^2$ with the mean energy in supernovae ejecta per unit mass formed, assuming a diet Salpeter IMF and an energy output of $10^{44}$~J per supernova event. The efficiency $\epsilon_{\halo}$ with which disk gas is reheated is a free parameter that is constrained by observations. If this energy exceeds that required for the heating, a fraction of cold gas will be ejected outside the halo,
\begin{equation}
  \Delta M_{\ejected} = \frac{2 \left( \Delta E_{\SN} - \Delta E_{\hot} \right)}{V_{\vir}^2} .
  \label{eqn:dMejected}
\end{equation}
The ejected gas is returned within a few halo dynamical times,
\begin{equation}
  \Delta M_{\returned} = \epsilon_{\returned} \frac{M_{\ejected} \Delta t}{t_{\dynHalo}} ,
  \label{eqn:dMreturned}
\end{equation}
where $t_{\dynHalo}=\frac{R_{\vir}}{V_{\vir}}$ and $\epsilon_{\returned} \approx 0.5$. If the available energy is less than that required for heating, no gas is ejected, and the reheated mass is
\begin{equation}
  \Delta M_{\heated} = \frac{2 \Delta E_{\SN}}{V_{\vir}^2} .
  \label{eqn:dMheated}
\end{equation}
It is worth noting that the exact values for most parameters used in this implementation are not important, as the supernovae feedback mechanism is parametrised by the quantity $\epsilon_{\halo}$, the magnitude of which is constrained by observations.

\subsection{Radio source feedback}
\label{sec:AGNfeedbackCh6}

Apart from forming stars, a fraction of the cold gas will be accreted onto a disk in the vicinity of the black hole. In order for the accreted gas to spiral in towards the centre, conservation of momentum requires its angular momentum be removed. Although magnetic fields could facilitate this process, in practice this requires ejection of gas. The power of the resultant jets depends on the amount of fuel available. When powerful enough, these jets will terminate in hotspots, and backflow of radio plasma will inflate the cocoons observed in powerful FR-II radio sources. The radio source will then expand, driving a shock through the intracluster gas and thereby heating the swept-up gas located between the contact discontinuity delineating the cocoon and the bow-shock. This heating will in turn limit the rate of gas cooling and subsequently the rate at which the black hole is fuelled, until production of jets of sufficient power to affect the gas is shut off. Once the gas has time to cool, the process is restarted. Below we develop a detailed model for following the interplay between the heating and cooling of the gas.

\subsubsection{Shock heating}
\label{sec:AGNheating}

Observations (e.g. Leahy \etal\/ 1989; Subrahmanyan \etal\/ 1996) suggest radio sources are self-similar, meaning the bow-shock and cocoon radii are related by $R_{\cocoon} = \lambda R_{\shock}$. We adopt the models of Kaiser \& Alexander (1997) and Alexander (2000, 2006) to describe the evolution of the cocoon radius $R_{\cocoon}$ with time. The gas density profile in each halo is approximated as a flat core~$-$ double power law atmosphere. In other words, the source initially evolves in a flat atmosphere, followed by two power law profiles of the form $\rho (r) = \rhoCore\/ \left( \frac{r}{\rCore\/} \right)^{-\beta}$. This would correspond to an expansion through a high-density core, into the galaxy, and then the cluster gas.

Both analytical work (e.g. Kaiser \& Alexander 1997) and numerical simulations \citep{ReynoldsEA01} suggest the swept-up gas lying between the shock and cocoon radii is isobaric. Such an arrangement is facilitated by backflow of swept-up gas from the hotspot to the sides of the cocoon. We therefore follow Heinz \etal\/ (1998) and Alexander (2002) and assume the cocoon to be spherical.

We further assume the swept-up gas evolves isothermally, consistent with cluster observations. Following the treatment of Alexander (2002), the supersonic expansion of the radio source shocks this gas to a temperature
\begin{equation}
  T_{\shocked} = \frac{15}{16} \frac{3-\beta}{11-\beta} \left( \frac{\mu m_{\rm H}}{k_{\rm B}} \right) \frac{1}{\lambda^2} \dot{R}_{\cocoon}^2 ,
  \label{eqn:Tshocked}
\end{equation}
where $\lambda$ and $\beta$ are related by $1-\lambda^3 = \frac{15}{4(11-\beta)}$.

Important parameters governing the radio source expansion and hence the heating of the gas are the gas density profile, which is given by Equation~\ref{eqn:rhoDM}; cocoon axial ratio $\RT\/$, which is related to the jet opening angle and is fixed at $\RT\/=2$, a value appropriate for Cygnus A \citep{KA97,BegelmanCioffi89}; and jet power $Q_{\rm jet}$. Following the dynamical model of Kaiser \& Alexander (1997), the cocoon radius is

\begin{equation}
  R_{\cocoon} (t) = a_D \rCore\/ \left( \frac{t}{\tau} \right)^{3/(5-\beta)} ,
  \label{eqn:cocoonSizeKA97}
\end{equation}
where $\tau = \left( \frac{2 R_{\rm core}^5 \rho_{\rm core}}{Q_{\rm jet}} \right)^{1/3}$ is a convenient timescale; and the dimensionless constant $a_D$ is given by \citep{KA97}

\begin{equation}
  a_D = \left[ \frac {(\Gamma_{\rm x}+1) (\Gamma_{\rm c}-1) (5-\beta)^3} {18 \pi \left( 9 \left[ \Gamma_{\rm c} + (\Gamma_{\rm c}-1) R_{\rm T}^2 \right] -4-\beta \right)} \right]^{1/(5-\beta)} \nonumber
  \label{eqn:cocoonSizeCoeffKA97}
\end{equation}
The adiabatic indices for the cocoon and external gas, $\Gamma_{\rm c}$ and $\Gamma_{\rm x}$ respectively, are equal to $5/3$, corresponding to non-relativistic material.

\subsubsection{Jet power}
\label{sec:jetPower}

Theoretical work (Narayan \etal\/ 1998; Meier 2001) as well as observations of black hole X-ray binaries (K\"ording \etal\/ 2006; Fender \etal\/ 2004) suggest that in both these objects and AGNs at least two different accretion states exist. In both cases angular momentum is removed from the accretion disk by viscosity, allowing the gas to spiral in towards the central black hole. At high inflow rates (compared to the Eddington rate), the accretion flow is described by a standard thin disk solution \citep{ShakuraSunyaev73}. Here, the accretion disk is geometrically thin and optically thick, and produces a quasi-blackbody spectrum. This flow is radiatively efficient, as all the energy released through viscous dissipation can be radiated away \citep{Narayan02}. By contrast, at low accretion rates the flow is geometrically thick and optically thin. As a result, the cooling times are long, and instead of being radiated away the thermal energy of the inflowing gas is advected inward. The resultant Advection Dominated Accretion Flow (ADAF; Narayan \& Yi 1995) is thus radiatively inefficient, and can produce powerful jets. 

Meier (2001) gives the jet power generated by a geometrically thick, optically thin ADAF disk as
\begin{eqnarray}
	Q_{\jetADAF} & = & 1.3 \times 10^{38} \left( \frac{\alpha_{\rm ADAF}}{0.3} \right)^{-1} g_{\rm ADAF}^2 (0.55 f_{\rm ADAF}^2 + 1.5 f_{\rm ADAF} j_{\BH\/} + j_{\BH\/}^2) \nonumber \\
	& & \times \left( \frac{M_{\BH\/}}{10^9 \Msun\/} \right) \left( \frac{\dot{M}_{\BH\/}}{\epsilon_{\rad\/} \dot{M}_{\edd}} \right)~{\rm W} 
	\label{eqn:QjetADAFgeneral}
\end{eqnarray}
for $\dot{m}_{\BH} \equiv \frac{\dot{M}_{\BH\/}}{\dot{M}_{\edd\/}} < \dot{m}_{\crit\/}$, where $\dot{m}_{\crit\/}$ is a parameter discussed in the following section.

Here, $\dot{M}_{\BH}$ is the black hole accretion rate, and $\dot{M}_{\edd\/}=\frac{L_{\edd\/}}{\epsilon_{\rad\/} c^2}=\frac{2.3}{\epsilon_{\rad\/}} \left( \frac{M_{\BH\/}}{10^8 \Msun\/} \right)$~\Msun\/\,yr$^{-1}$ is the Eddington accretion rate for a radiative efficiency $\epsilon_{\rad\/}$. Disk viscosity is responsible for transporting angular momentum outward, and is given by the coefficient $\alpha_{\rm ADAF} \sim 0.3$ \citep{NarayanEA98,Narayan02}. Coefficients $g_{\rm ADAF}$ and $f_{\rm ADAF}$ refer to the ratios of actual angular velocity and azimuthal magnetic field to those calculated by Narayan \& Yi (1995), and $j_{\BH\/}$ is the black hole spin. Following Meier (2001) and Okamoto \etal\/ (2008b) we adopt $g_{\rm ADAF}=2.3$, $f_{\rm ADAF}=1$ and $j_{\BH\/}=0.5$, noting that the jet power is not very sensitive to realistic changes in these values. For this choice of parameters, Equation~\ref{eqn:QjetADAFgeneral} can be rewritten as
\begin{eqnarray}
	Q_{\jetADAF} & = &4.5 \times 10^{39} \left( \frac{\dot{M}_{\BH}}{\Msun\/\,{\rm yr^{-1}}} \right)~{\rm W} \nonumber \\
	& = & 0.79 \dot{M}_{\BH} c^2 .
	\label{eqn:QjetADAF}
\end{eqnarray}
For accretion rates $\dot{M}_{\BH} > \dot{m}_{\crit} \dot{M}_{\edd}$ the accretion flow is described by the standard Shakura-Sunyaev thin disk solution, which for the adopted viscosity parameter is given by
\begin{equation}
	Q_{\jetTD} = 6.0 \times 10^{-4} \left( \frac{M_{\BH}}{10^9 \Msun\/} \right)^{-0.3} \left( \frac{\dot{M}_{\BH}}{\Msun\/\,{\rm yr^{-1}}} \right)^{0.2} \dot{M}_{\BH} c^2 .
	\label{eqn:QjetTD}
\end{equation}
This mode of accretion (referred to as the ``quasar mode'' by Croton \etal\/ 2006) is radiatively efficient, and typically results in weaker jets than the ADAF case. Thus it is identified with a luminous AGN phase \citep{CrotonEA06,BowerEA06}; while the ADAF phase corresponds to the ``radio mode'' of Croton \etal\/ and is responsible for radio source feedback. In what follows, we assume that when the {\itshape instantaneous} black hole accretion rate exceeds the critical rate, the disk is in the ``quasar mode''; and when the instantaneous accretion rate is below this value, a radio jet with power given by Equation~\ref{eqn:QjetADAF} is generated. We further assume that in this case the {\itshape actual} accretion rate onto the black hole is given by $\dot{M}_{\BH\/} = \dot{m}_{\crit\/} \dot{M}_{\edd\/}$, i.e. the disk is fuelled at the maximum rate (e.g. Hopkins \etal\/ 2006). The usual radiative efficiency of $\epsilon_{\rad\/}=0.1$ is assumed in evaluating $\dot{M}_{\edd\/}$.

\subsubsection{Radio source intermittency}
\label{sec:FRIvsFRII}

Croton \etal\/ (2006) and Bower \etal\/ (2006) both took the ``radio mode'' heating provided by the jets to be continuous. Observations of black hole binaries and the apparent balance between heating and cooling rates, however, suggest otherwise. Jets will be intermittent for two reasons. Feedback limits the availability of cold gas required to fuel the central engine \citep{ShabalaEA08}, resulting in weaker jets that are more liable to disruption by instabilities in the dense central regions. On the other hand, when the radio cocoon expands outside the high density regions, the central gas cools rapidly. The resulting high black hole accretion rates qualitatively change the nature of the accretion flow from a radiatively inefficient ADAF to a radiatively efficient standard thin disk flow \citep{NarayanEA98}. This once again results in a much weaker jet (by $\sim 3$ orders of magnitude according to Equations~\ref{eqn:QjetADAF} and \ref{eqn:QjetTD}) which can be disrupted.

In the assumed density profile, the jets are most likely to only be susceptible to disruption by Kelvin-Helmholtz instabilities in the denser central regions of the halo \citep{Alexander02,KA97}. Following the analysis of Alexander (2002), the minimum power a jet must have to traverse the flat core is taken as
\begin{equation}
	Q_{\jetMin\/} = 10^{37} \left( \frac{R_{\core\/}}{2.5\,{\rm kpc}} \right)^2 \left( \frac{\rho_{\core\/}}{1.5 \times 10^{22}~{\rm kg\,m^{-3}}} \right)~{\rm W} .
	\label{eqn:QjetMinStable}
\end{equation}

Similarly, for a radio source that has expanded outside $R_{\core\/}$, the jet power must exceed $Q_{\jetMin\/} \propto R_{\cocoon\/}^{2-\beta} \rho_{\core\/}$ in order to reach the hotspots. If this does not happen, or the instantaneous rate of accretion onto the black hole $\dot{m}_{\BH\/}$ exceeds $\dot{m}_{\crit\/}$ causing the disk to become radiatively inefficient, the jet is terminated, and a new jet inflates a new radio cocoon. For a radio source that is still within the core, i.e. when $R_{\cocoon\/}<R_{\core\/}$, the minimum required jet power is given by replacing the core radius with the cocoon radius in Equation~\ref{eqn:QjetMinStable}.

The final important point concerning jet intermittency relates to the critical accretion rate denoting the transition between radiatively efficient and inefficient disks. Narayan \etal\/ (1998) estimate that an ADAF will switch to a radiatively efficient (thin disk) flow when the accretion rate in Eddington units exceeds $\dot{m}_{\critUp\/} \sim \alpha_{\rm ADAF}^2 \approx 0.1$ for the adopted viscosity value. This threshold is consistent with the observed transition from low/hard to high/soft state in X-ray binaries \citep{EsinEA98,KoerdingEA06}. Interestingly, the reverse transition is less well defined, with $\dot{m}_{\critDown\/} \sim 0.01 - 0.1$ \citep{Narayan02}. The thin disk will develop a hole in its inner regions, where an ADAF is allowed. As the accretion rate drops further, the region hosting an ADAF will grow, until this becomes the dominant mode of accretion. Hence, in the present model we treat $\dot{m}_{\critDown\/}$ as a free parameter in the range $0.01-0.1$.

\section{Constraints on model parameters}
\label{sec:parameterSpace}

There are three free parameters in the presented model. Supernovae heating is parametrised by $\epsilon_{\halo\/}$, the efficiency with which supernovae heat the cold disk gas. The other two parameters relate to AGN feedback. Fraction of cold disk gas accreted onto the black hole is given by $\epsilon_{\acc\/}$, and $\dot{m}_{\critDown\/}$ denotes the dimensionless accretion rate (in Eddington units) below which the accretion flow switches from a thin disk to an ADAF solution.

\begin{figure*}[tbph]
\centering
  \subfigure[$1.6 \times 10^{11}$~\Msun\/]{\includegraphics[height=0.45\textwidth,angle=-90]{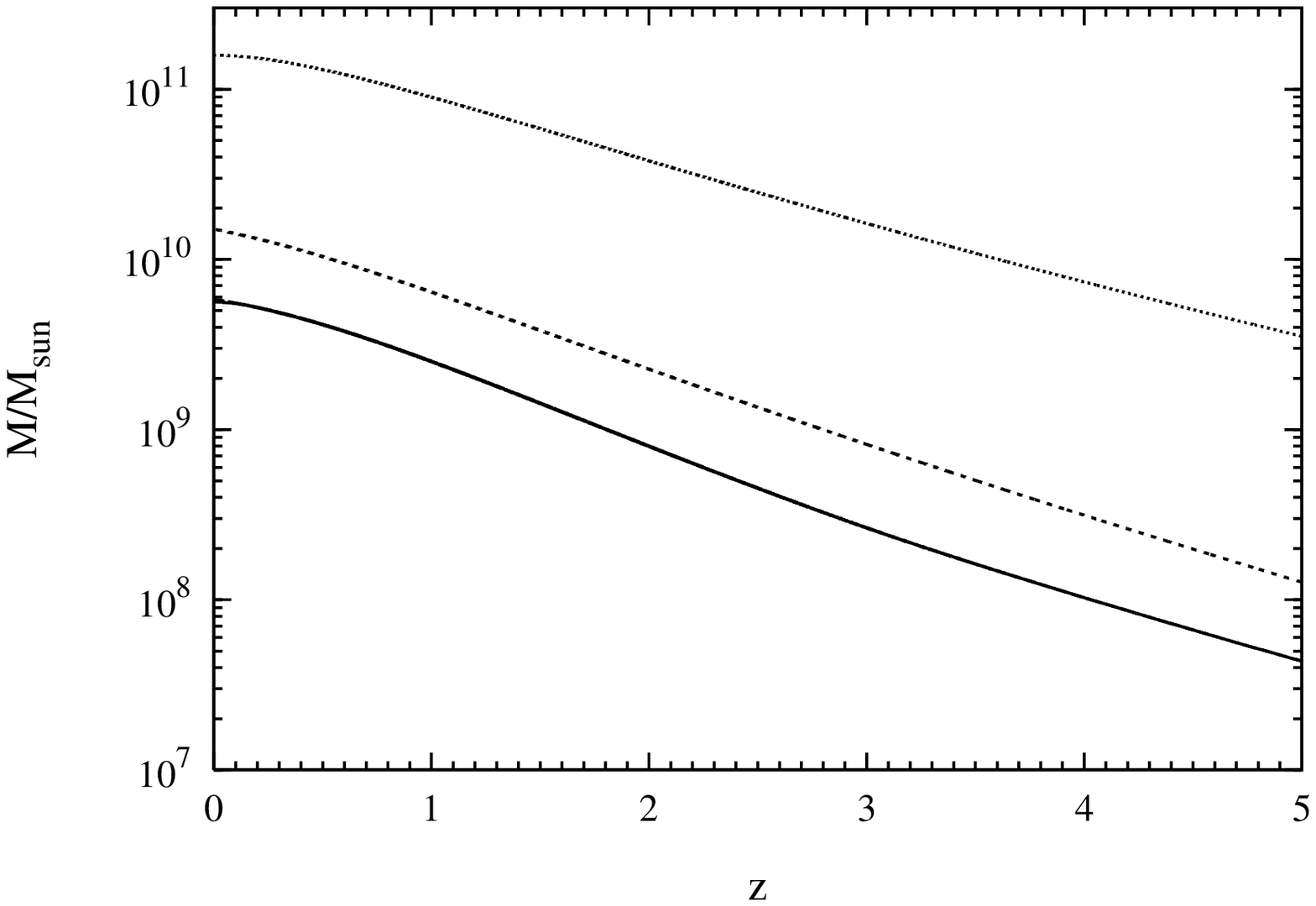}}
  \subfigure[$1.6 \times 10^{15}$~\Msun\/]{\includegraphics[height=0.45\textwidth,angle=-90]{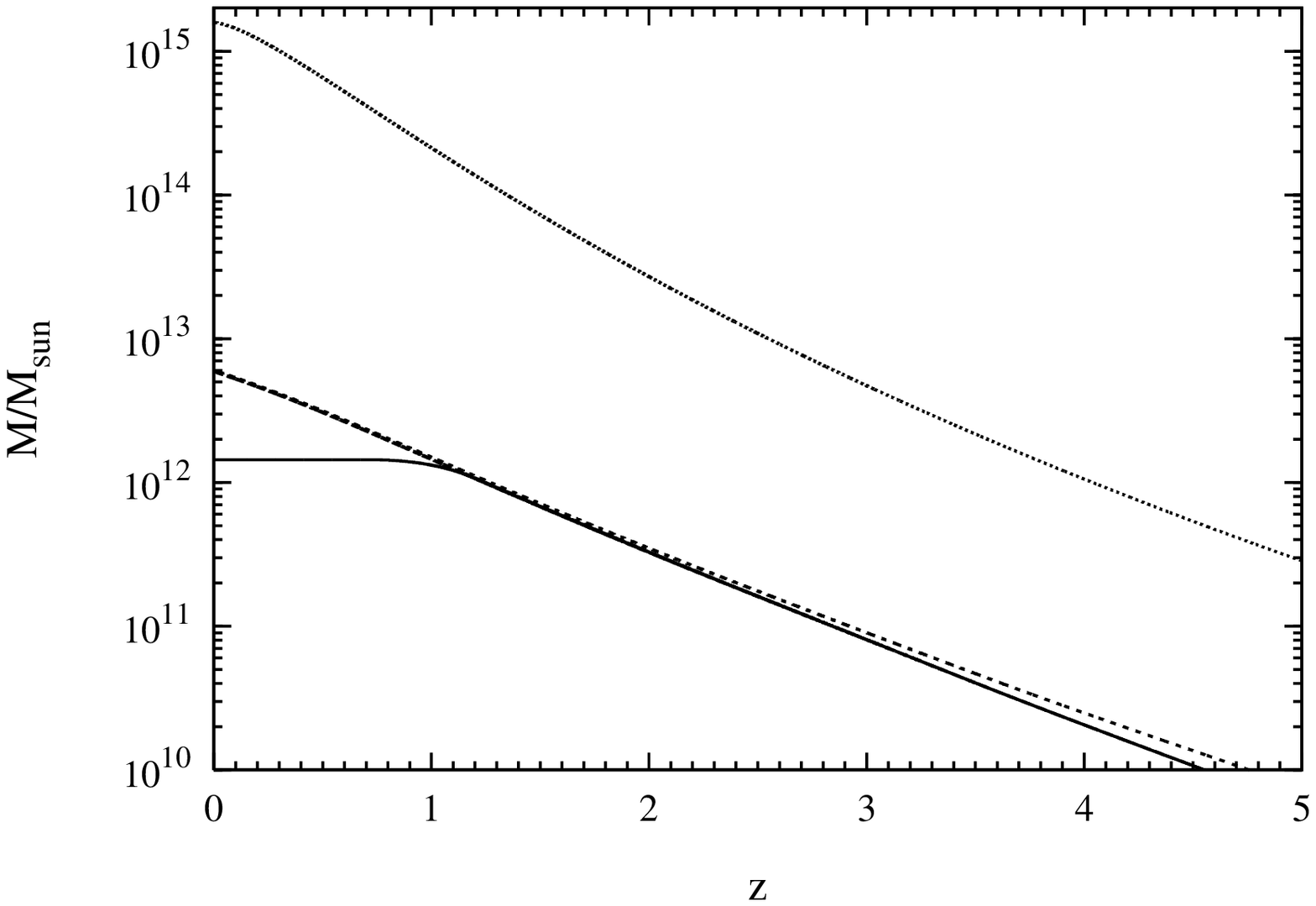}}
\caption{Evolution of the stellar mass content in haloes with redshift zero masses of $1.6 \times 10^{11}$ and $1.6 \times 10^{15}$~\Msun\/. Different curves correspond to different models: no feedback (short-dashed); reionization and supernovae heating (long-dashed); and full model including AGN feedback (solid line). Halo mass build-up (dotted line) is plotted for reference. All curves are for $\epsilon_{\halo\/}=0.01$, $\epsilon_{\acc\/}=0.02$, $\dot{m}_{\critDown\/}=0.03$. Reionization and SNe feedback are important at early times and in low-mass haloes, while radio source feedback dominates at late epochs in massive hosts. Thus, on scales of interest the reionization and supernovae heating (long-dashed) curve coincides with the full model (solid) curve for the $1.6 \times 10^{11}$~\Msun\/ halo; and the no feedback (short-dashed) and reionization plus supernovae (long-dashed) curves coincide for the $1.6 \times 10^{15}$~\Msun\/ halo.}
\label{fig:paper_parameterSpace_differentPhases}
\end{figure*}

Constraints on these parameters come from observations. Radio source heating is only important in massive haloes and at late times, where the gas cooling rate can provide sufficient fuel for a radio source. This can be seen in Figure~\ref{fig:paper_parameterSpace_differentPhases}, where evolution of the stellar component is shown for haloes with masses $10^{11}$ and $10^{15}$~\Msun\/ at the present epoch. In the $10^{11}$~\Msun\/ halo, supernovae feedback and reionization heating result in the stellar content of the halo being at all times consistently lower by a similar factor than when only gas cooling is included. By contrast, this has little effect on the $10^{15}$~\Msun\/ halo, and even then only at the highest ($z>3$) redshifts. However, at $z \sim 1$ the cooling rate becomes high enough for radio source feedback to become important. From that point it is AGN feedback that dominates the cooling history of the halo.

\begin{figure}[tbph]
\centering
  \includegraphics[height=0.65\textwidth,angle=-90]{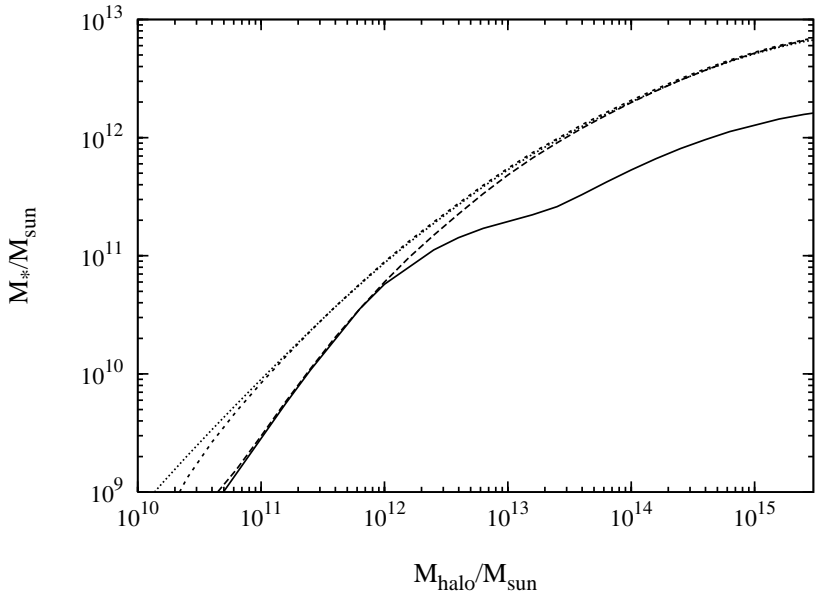}
\caption{Stellar mass at the present epoch for various models: cooling (dotted); reionization (short-dashed); reionization and supernovae (long-dashed); full feedback model including AGN (solid line). The curves are again for $\epsilon_{\halo\/}=0.01$, $\epsilon_{\acc\/}=0.02$, $\dot{m}_{\critDown\/}=0.03$. AGN feedback is important in massive ($M_{\halo}>10^{12}$~\Msun\/) galaxies.}
\label{fig:paper_MstarsMhalo}
\end{figure}

Contributions of individual feedback components to halting the build-up of stellar mass are shown in Figure~\ref{fig:paper_MstarsMhalo}. Feedback from supernovae and reionization is dominant in low-mass ($\leq 10^{12}$~\Msun\/) haloes, while star formation is significantly impaired by AGN feedback in massive haloes.

\subsection{Initial Mass Function}
\label{sec:derivedQuantities}

Both the rest-frame UV flux used to derive star-formation rates, and optical and infra-red emission tracking the integrated stellar mass are heavily obscured by dust \citep{SteidelEA99,MadauEA98,GiavaliscoEA04}. Fortunately, the rest-frame near infra-red {\it K}-band luminosity function is almost a magnitude less sensitive to dust extinction than at other wavelengths \citep{KauffmannEA99a}, making it a very useful probe of stellar mass.

The main uncertainty in converting between stellar masses and luminosities therefore comes from the adopted IMF. The standard relation frequently used is the Salpeter (1955) IMF given by $\frac{d N}{d \log M} \propto M^{-1.35}$ derived for solar neighbourhood stars. It has recently been argued, however, that field stars do not provide a good proxy for the low-mass end of the IMF \citep{Scalo98} due to the sensitivity of the derived mass function to star formation histories at these masses, as well as being at odds with H\,$\alpha$ luminosities of galaxy populations \citep{BaldryEA04}. Studies of star clusters and galaxies suggest that a ``universal'' IMF (in a statistical sense) may be applicable \citep{Scalo98}. To this end, a number of models have been put forward. Kennicutt (1983), Kroupa (2001) and Chabrier (2003) IMFs are in good agreement with the Salpeter one at high masses, but predict less low-mass stars. Bell \etal\/ (2003) adopt a ``diet Salpeter'' IMF, which has the same colours and luminosities as the normal Salpeter IMF, but only 70\% of the mass due to a lower number of low-mass (and therefore faint) stars in order to match observed rotation velocities in the Ursa Major cluster. Using the vertical velocity dispersion of stars in disk-dominated galaxies, Bottema (1997) argued for an even lower mass-to-light ratio. As a result, the uncertainty in the IMF slope is as much as 0.5 \citep{BaldryEA04}.

Bruzual \& Charlot (2003) compared the spectral properties of a number of IMFs, and found that the (${\it B-V}$) and (${\it V-K}$) colours predicted by these are very similar. The main differences arise in the mass-to-light ratios. Bell \etal\/ (2003) estimate that the choice of IMF results in the mass-to-light ratio varying by up to $0.5$~dex, with top-heavy IMFs having lower mass-to-light ratios.

\subsection{Efficiency of supernovae feedback}
\label{sec:effHalo}

\subsubsection{Counts of faint galaxies}
\label{sec:effHalo_LFandMF}

As Figures~\ref{fig:paper_parameterSpace_differentPhases} and \ref{fig:paper_MstarsMhalo} show, low-mass haloes are little affected by AGN feedback. Since these haloes host galaxies with relatively low stellar masses, the low-mass end of the stellar mass function (or, correspondingly, the faint end of the {\it K}-band luminosity function) can be used to constrain the contribution of supernovae to suppressing star formation.

\begin{figure}[tbph]
\centering
  \includegraphics[height=0.65\textwidth,angle=-90]{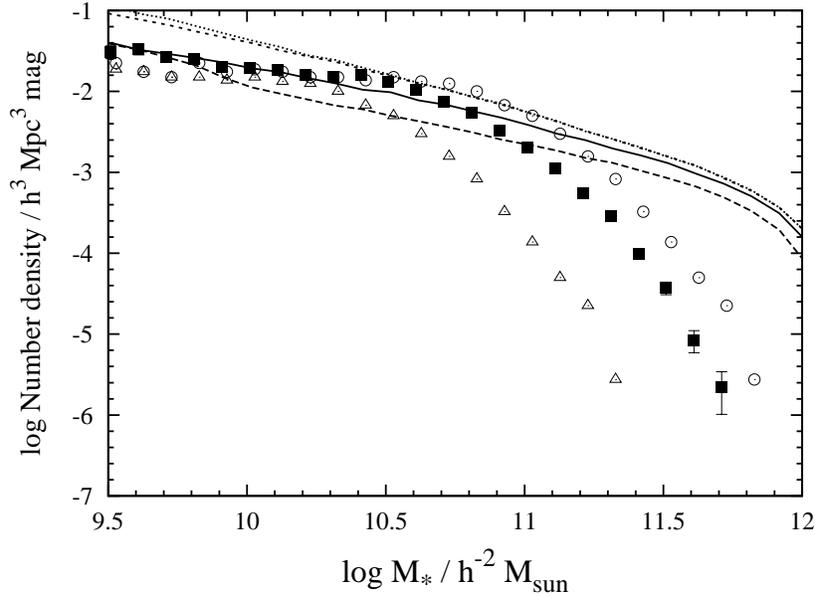}
\caption{Local stellar mass function for various supernovae feedback efficiencies. Observed mass functions are derived from the 2dFGRS {\it J}-band near infra-red observations \citep{ColeEA01}, assuming a Salpeter (open circles), Bottema (open triangles) or diet Salpeter (filled squares) IMF. Model curves are: cooling (dotted), reionization (short-dashed), reionization plus supernovae feedback with $\epsilon_{\halo} = 0.01$ (solid), $\epsilon_{\halo} = 0.04$ (long-dashed). Bright end counts are significantly overpredicted without AGN feedback. A colour version of this Figure is available in the online article.}
\label{fig:paper_constrain_effHalo_LFandMF}
\end{figure}

In Figure~\ref{fig:paper_constrain_effHalo_LFandMF} the local stellar mass function is plotted for various values of $\epsilon_{\halo}$. The curves are constructed by convolving the star formation histories of individual haloes with the halo mass function obtained at each redshift from the Millennium Simulation (Section~\ref{sec:haloEvolution}). Shown by open symbols are the stellar mass functions derived from the Cole \etal\/ (2001) 2dFGRS {\itshape K}-band luminosity function assuming Salpeter and Bottema IMFs. These correspond to the upper and lower limits, respectively, on the mean mass-to-light ratio used to convert from luminosities to stellar masses \citep{BellEA03}. In the case of Salpeter IMF, we use the average mass-to-light ratio of 1.32 as quoted by Cole \etal\/ For the Bottema IMF this value is reduced to 0.82. The filled symbols correspond to the stellar mass function derived with a diet Salpeter IMF, obtained by decreasing the Salpeter mass-to-light ratio by 0.15 dex to 1.17 \citep{BellEA03}; this gives good agreement with the Cole \etal\/ stellar mass function derived from {\itshape J}-band observations. 

It can be seen that $\epsilon_{\halo}=0.01$ provides a good match to the local stellar mass function at the low-mass end. Although the choice of IMF impacts the derived mass function at the bright end, in the absence of AGN feedback the number densities of the most massive galaxies are overpredicted significantly.

\subsubsection{Distribution of stellar mass}
\label{sec:effHalo_SFR}

\begin{figure}[tbph]
\centering
  \includegraphics[height=0.65\textwidth,angle=-90]{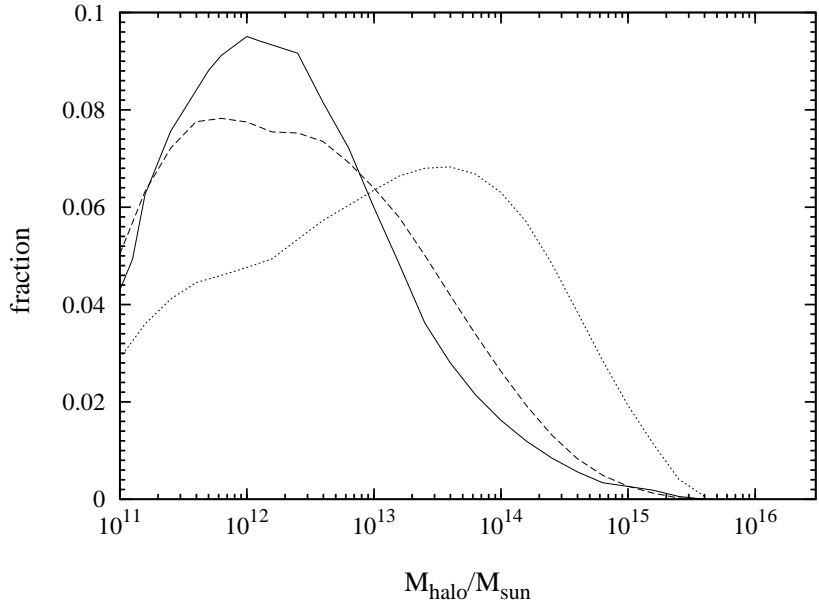}
\caption{Fraction of stellar mass contained in different mass haloes at the present epoch. Dashed curve represents the cooling-only model. Solid curve includes reionization heating, supernovae feedback for $\epsilon_{\halo}=0.01$, and AGN feedback with $\epsilon_{\acc}=0.02$ and $\dot{m}_{\critDown}=0.03$. Fractional distribution of halo mass (dotted curve) is plotted for comparison.}
\label{fig:paper_fracMstarsMhalo}
\end{figure}

Low mass haloes are much more abundant than their higher mass counterparts- for example, $10^{12}$~\Msun\/ haloes are five orders of magnitude more abundant than $10^{15}$~\Msun\/ ones \citep{JenkinsEA01}. Since massive haloes host the more massive galaxies, however, the question of how the total stellar mass is distributed between haloes is a non-trivial one. We plot in Figure~\ref{fig:paper_fracMstarsMhalo} the stellar mass fraction as a function of halo mass at the present epoch. Haloes with masses less than $3 \times 10^{11} - 3 \times 10^{12}~\Msun\/$ are seen to dominate the total stellar mass budget, despite the total mass distribution peaking at higher halo masses.

\begin{figure}[tbph]
\centering
  \includegraphics[height=0.65\textwidth,angle=-90]{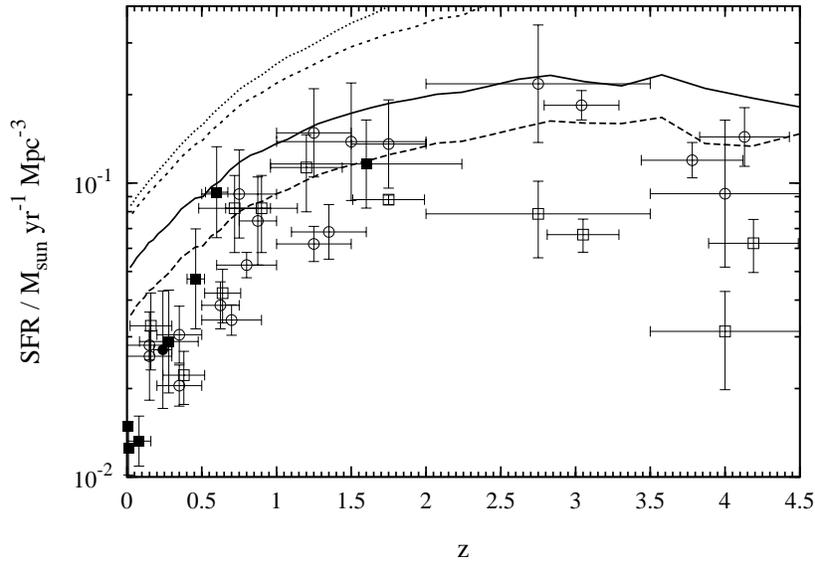}
\caption{Evolution of the global star formation rate density. Various curves denote models with: cooling only (dotted); reionization (short-dashed); and reionization plus supernovae with $\epsilon_{\halo}=0.01$ (solid) and $\epsilon_{\halo}=0.04$ (long-dashed). Observations are from the compilation of Hopkins (2004) and {\it J}-band observations of Cole \etal\/ (2001; open squares), assuming a diet Salpeter IMF. Hopkins (2004) points represent various UV (open circles), radio (filled squares) and X-ray (filled circles) observations. A colour version of this Figure is available in the online article.}
\label{fig:paper_constrain_effHalo_SFR}
\end{figure}

This result suggests that the star formation history of the Universe is largely determined by reionization and supernovae, rather than AGN, feedback. This is certainly true for $z>1$, when AGN feedback does not greatly affect the evolution of the stellar mass content even in the most massive haloes (Figure~\ref{fig:paper_parameterSpace_differentPhases}). Hence a guide to the efficiency of the SNe feedback can be obtained from observations of the evolution of the star formation rate (SFR) density, plotted in Figure~\ref{fig:paper_constrain_effHalo_SFR}. Observations at $z>1$ suggest $\epsilon_{\halo} \approx 0.01$. It is also clear from Figure~\ref{fig:paper_constrain_effHalo_SFR} that models without AGN feedback significantly overpredict star formation rates at later epochs. The observed star formation rates are derived from a combination of UV, IR, O\,{\small [II]}, radio and X-ray observations (see caption) assuming a diet Salpeter IMF. It is worth noting that, due to the assumed IMF, these are close to the upper limits on the actual star-formation rates at each epoch, and the actual SFRs can be up to 0.35 dex lower than these values.

\subsection{AGN fuelling efficiency}
\label{sec:effAcc}

With supernovae feedback efficiency fixed, the fraction of gas accreted onto the central black hole can be determined by invoking known observed relations between the mass of the central black hole and global halo and galaxy properties. One of these is the correlation between the local masses of the black hole and stars in the bulge (as derived from their luminosities; Magorrian \etal\/ 1998; H\"aring \& Rix 2004).

\begin{figure}[tbph]
\centering
  \includegraphics[height=0.65\textwidth,angle=-90]{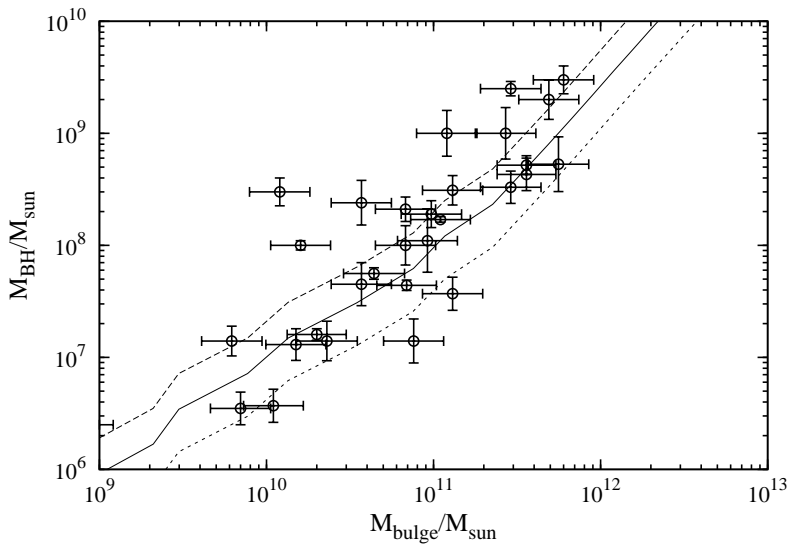}
\caption{$M_{\BH}-M_{\bulge}$ relation in the local Universe as a function of accretion efficiency. Curves are: $\epsilon_{\acc}=0.01$ (short-dashed); $\epsilon_{\acc}=0.02$ (solid); and $\epsilon_{\acc}=0.05$ (long-dashed line). All models include reionization and supernovae feedback with $\epsilon_{\halo}=0.01$, but no AGN feedback. Data points are from H\"aring \& Rix (2004).}
\label{fig:paper_constrain_effAcc_MbhMbulge}
\end{figure}

Although an estimate of the stellar content in the bulge can be obtained by invoking disk stability arguments \citep{EfstathiouEA82}, such a formulation ignores mergers and thus attributes too much stellar mass to the disk. Benson \etal\/ (2007) determined spheroid-to-disk light ratios in a sample of $\sim 9000$ SDSS galaxies, and found that $\frac{M_{\bulge}}{M_{\star}} \sim 0.3 - 1.0$ for galaxies with stellar masses between $10^9$ and $10^{12}$~\Msun\/. We adopt the Benson \etal\/ relation between stellar mass and the fraction of stars located in the spheroid. Figure~\ref{fig:paper_constrain_effAcc_MbhMbulge} plots the $M_{\BH}-M_{\bulge}$ relation thus determined for models without any AGN feedback. Accreted fractions $\epsilon_{\acc} \sim 0.02-0.05$ are required to match observations of low-mass ($M_{\bulge}<10^{11}$~\Msun\/) galaxies.

Because our model does not account for galaxy groups and clusters, here the model bulge masses refer to the sum of stellar masses of all the galaxies in a given halo, multiplied by the bulge fraction. Although this fraction is likely to vary between objects (in particular, satellites are more likely to be disk-dominated than cD galaxies), one would expect such bulge estimates to be reasonably accurate because of the dominant contribution of the central galaxy to the stellar mass budget.

Due to the self-regulatory nature of the feedback processes, even in the presence of AGN feedback the nature of the $M_{\BH}-M_{\bulge}$ relation at the high-mass end is not expected to be altered substantially. Rather, one would expect both the black hole and bulge mass to be reduced by a similar factor relative to the no-feedback case. This point is considered in Section~\ref{sec:MbhMbulge}.

\subsection{Thin disk $-$ ADAF transition}
\label{sec:mdotCritDown}

The dimensionless accretion rate $\dot{m}_{\critDown}$ regulates how often jets are re-triggered. An observational constraint on the efficiency of feedback is provided by the bright end of the optical luminosity (or stellar mass) function.

\begin{figure}[tbph]
\centering
  \includegraphics[height=0.65\textwidth,angle=270]{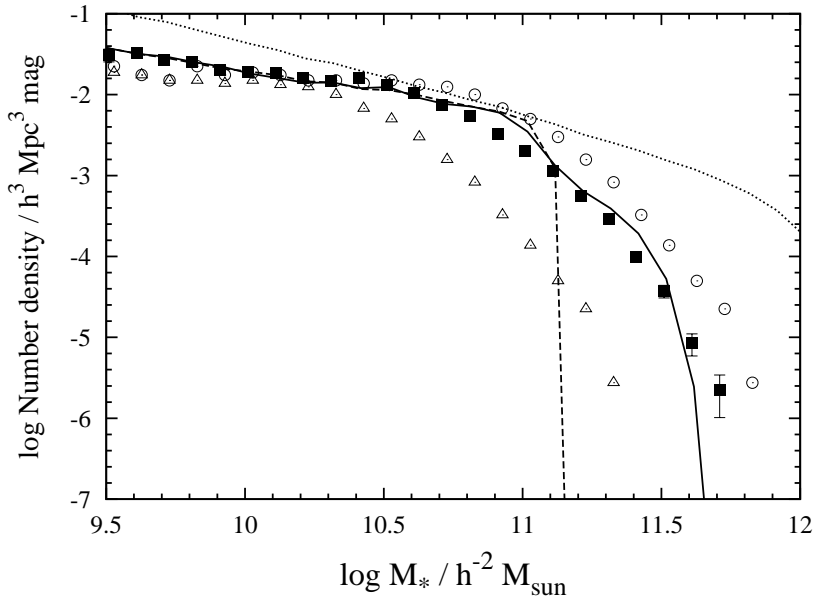}
\caption{Local stellar mass function for various $\dot{m}_{\critDown}$ values. Reionization and supernovae feedback for $\epsilon_{\halo}=0.01$ are included. Curves are for: $\dot{m}_{\critDown}=0.03$ (solid), $\dot{m}_{\critDown}=0.07$ (dashed). The no-AGN prediction (dotted curve) is plotted for comparison. Observed mass functions are as in Figure~\ref{fig:paper_constrain_effHalo_LFandMF}. A colour version of this Figure is available in the online article.}
\label{fig:paper_constrain_mdotDown_LFandMF}
\end{figure}

Figure~\ref{fig:paper_constrain_mdotDown_LFandMF} shows the $z=0$ stellar mass function for a diet Salpeter IMF for a range of $\dot{m}_{\critDown}$ values. The best fit is given by $\dot{m}_{\critDown} \approx 0.4$. Importantly, this set of plots shows that realistic AGN feedback can resolve the number counts problem at the high-mass end of the stellar mass function by suppressing star formation in massive haloes.

\subsection{AGN feedback models}
\label{sec:intermittentVsCroton}

\begin{figure*}[tbph]
\centering
  \subfigure[]{\includegraphics[height=0.32\textwidth,angle=270]{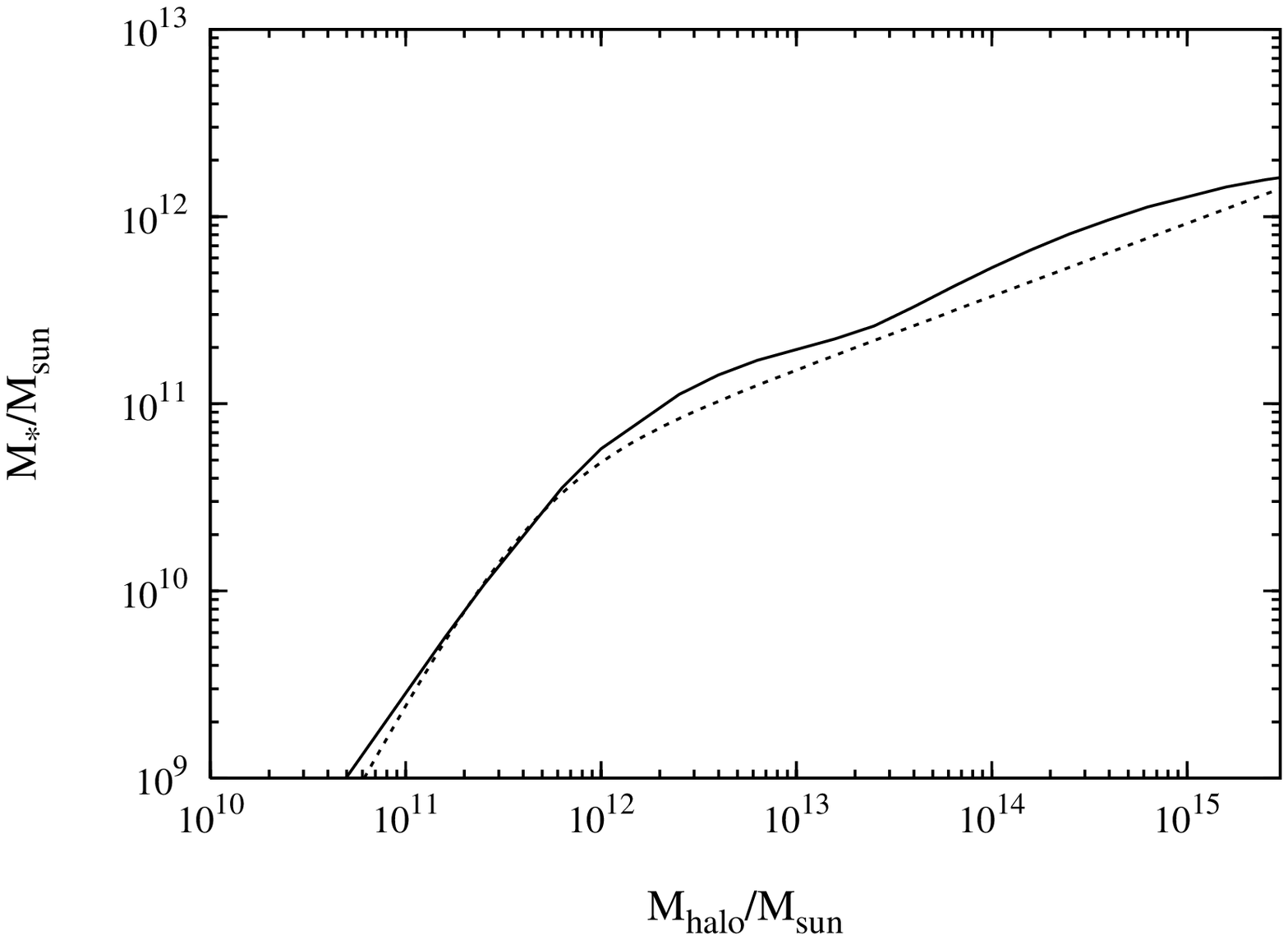}}
  \subfigure[]{\includegraphics[height=0.32\textwidth,angle=270]{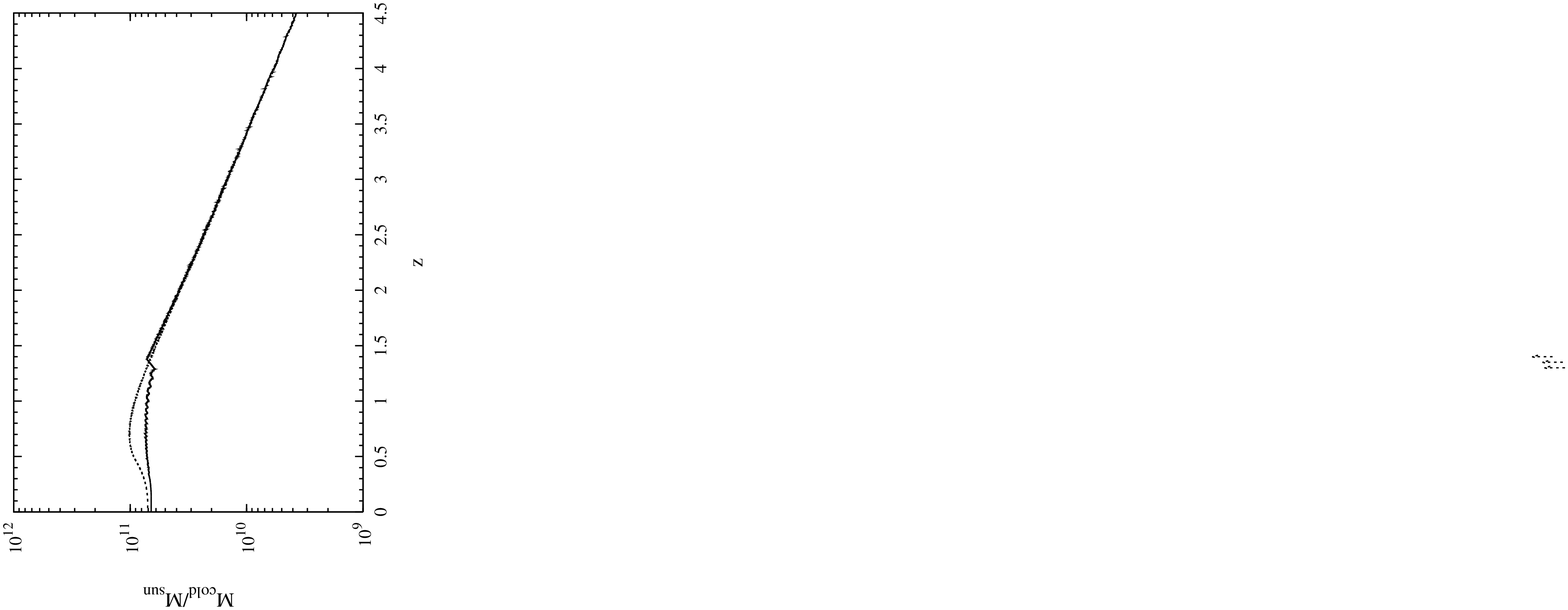}}
  \subfigure[]{\includegraphics[height=0.32\textwidth,angle=270]{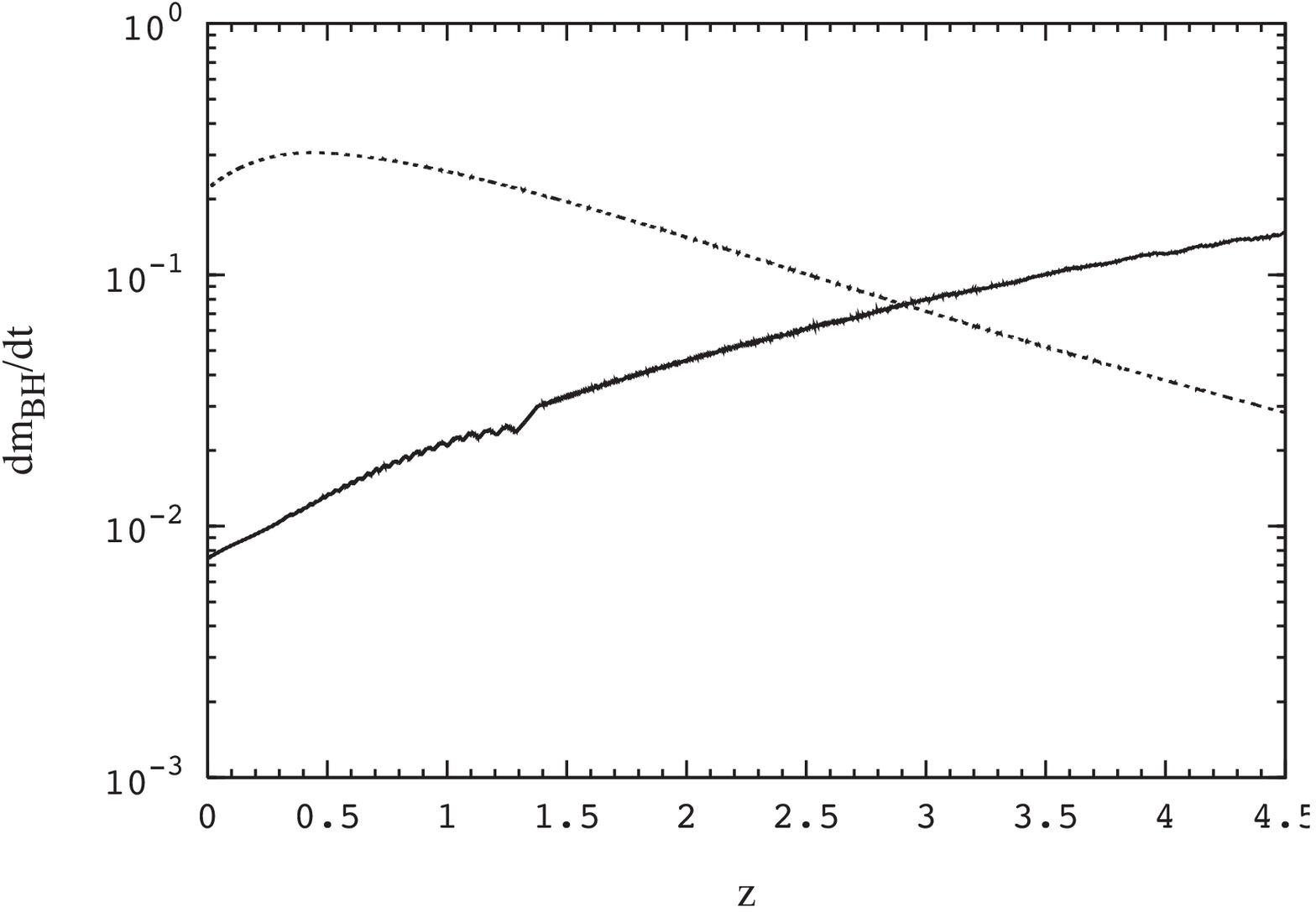}}
\caption{Comparison of AGN feedback models. {\it Left panel}: Stellar masses at $z=0$. Solid curve represents our model with the same parameters as in Figure~\ref{fig:paper_MstarsMhalo}. Results from Wang \etal\/ (2006) from the Millennium Simulation, with feedback as implemented by Croton \etal\/ (2006), are shown by a dotted curve. {\it Middle panel}: Cooling histories for a $10^{13}$~\Msun\/ halo. {\it Right panel}: Accretion rate onto the black hole, in Eddington units. We predict lower rates of accretion than the model with continuous feedback and $\dot{m}_{\BH}$ decreases at lower redshifts. The epoch of powerful AGN activity begins at $z \sim 1.4$ as $\dot{m}_{\BH}$ falls below $\dot{m}_{\critDown}=0.03$.}
\label{fig:paper_compare_intermittentVsCroton}
\end{figure*}

A number of authors (e.g. Croton \etal\/ 2006; Bower \etal\/ 2006; Cattaneo \etal\/ 2006; Granato \etal\/ 2004) have implemented AGN feedback in galaxy formation models. The major difference between their results and ours lies in the intermittency of our physically-motivated prescription. Figure~\ref{fig:paper_compare_intermittentVsCroton} compares our predictions to the model of Croton \etal\/ (2006). It is worth noting that the value of the scaling constant describing the strength of AGN feedback ($\kappa_{\rm AGN}$ in Croton \etal\/'s Equation 10) is different to the value obtained by those authors. This is a result of the differing model implementations.

Since both models reproduce the observed local stellar mass function, the stellar masses for each halo at the present epoch are in agreement (left panel). Cooling histories (middle panel) are also similar. However, the two models differ in their predictions of the rate at which cold gas is accreted onto the central black hole (right panel). The Croton \etal\/ model predicts an increasing (in Eddington units) accretion rate; while our model predicts $\dot{m}_{\BH}$ to decrease with time. At $z \sim 1.4$ the accretion rate falls for the first time below $\dot{m}_{\critDown}$, triggering a powerful jet episode and resultant intermittent activity. The interplay between gas heating, cooling and ejection by powerful AGN outbursts is crucial to explaining both the observed duty cycle of radio galaxies and quasars (Shabala \etal\/ 2008) and their evolution. We defer a detailed discussion of this point to a future paper.

\section{Cosmic downsizing and suppression of star formation}
\label{sec:results}

\subsection{$M_{\BH\/}$ $-$ $M_{\bulge\/}$ relation}
\label{sec:MbhMbulge}

\begin{figure}[tbph]
\centering
  \includegraphics[height=0.65\textwidth,angle=-90]{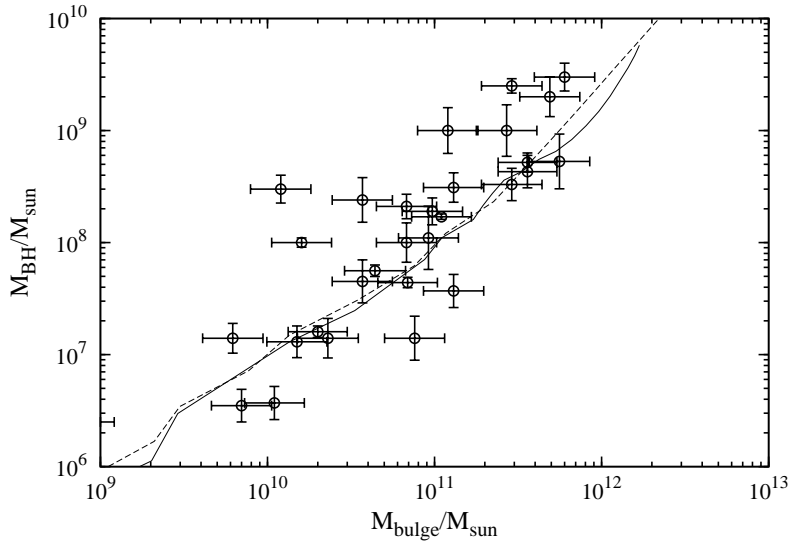}
\caption{$M_{\BH}-M_{\bulge}$ relation at z=0 for the best-fit ($\epsilon_{\halo}=0.01$, $\epsilon_{\acc}=0.02$ and $\dot{m}_{\crit}=0.03$) model (solid line). Data points are again from H\"aring \& Rix (2004). Prediction for the no-AGN case (dashed line) is plotted for comparison; this corresponds to the solid curve in Figure~\ref{fig:paper_constrain_effAcc_MbhMbulge}.}
\label{fig:paper_MbhVsMhalo}
\end{figure}

Figure~\ref{fig:paper_constrain_mdotDown_LFandMF} shows that the model can explain the observed dearth of massive galaxies and match the stellar mass function at the present epoch. It must also reproduce the observed local correlation between black hole mass and global properties such as spheroid mass. This is plotted in Figure~\ref{fig:paper_MbhVsMhalo}.

Although the final stellar and black hole masses in each halo are lower than when feedback is not included, these clearly grow ``in step'', with the slope of the final relation remaining unchanged. Thus radio source heating provides a self-regulatory mechanism that can simultaneously limit the growth of the central black hole and the bulge by a similar amount. 

\subsection{Evolution of global properties}
\label{sec:resultsVsz}

\subsubsection{Star formation rate}
\label{sec:SFRvsz}

\begin{figure}[tbph]
\centering
  \includegraphics[height=0.65\textwidth,angle=-90]{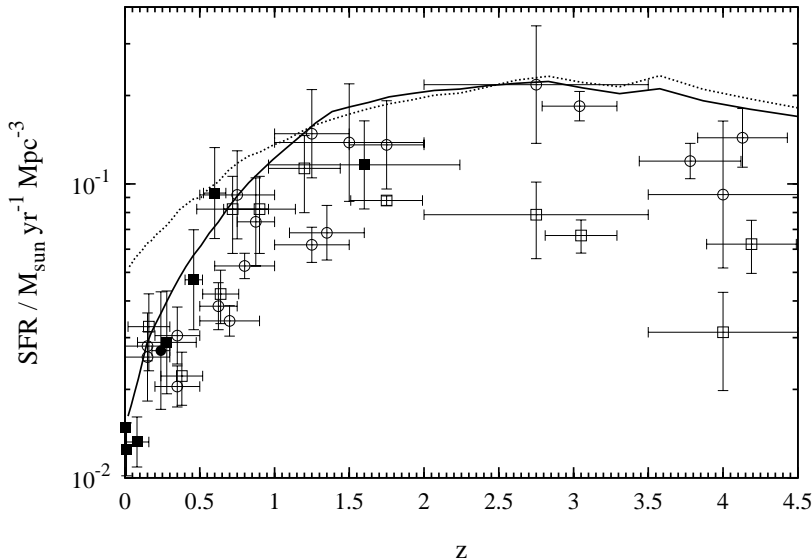}
\caption{Evolution of the global star formation rate density. Solid curve represents the best-fit model ($\epsilon_{\halo}=0.01$, $\epsilon_{\acc}=0.02$, $\dot{m}_{\critDown}=0.03$) as given by the local stellar mass function. Dotted curve is for the same parameters but without AGN feedback. Observed points are as in Figure~\ref{fig:paper_constrain_effHalo_SFR}.}
\label{fig:paper_SFR}
\end{figure}

Figure~\ref{fig:paper_SFR} plots the star formation rate density as a function of cosmic epoch for the model providing the best fit to the local stellar mass function (Figure~\ref{fig:paper_constrain_mdotDown_LFandMF}), $\epsilon_{\halo}=0.01$, $\epsilon_{\acc}=0.02$, $\dot{m}_{\critDown}=0.03$. The no-AGN predictions are also plotted for comparison. As argued in Section~\ref{sec:effHalo_SFR}, models without AGN feedback significantly overpredict the global star formation rate density at the present epoch. Inclusion of such feedback, on the other hand, provides results that are in good agreement with observations.

Integrating the predicted local stellar mass function yields a total present-day stellar mass density of $\rho_{\star} = 4.5$~\Msun\,Mpc$^{-3}$, or $\Omega_{\star} = 3.3 \times 10^{-3}$. This is consistent with observational estimates of $\Omega_{\star} = (2.8 \pm 0.4) \times 10^{-3}$ \citep{ColeEA01,BellEA03} and $(3.5 \pm 0.4) \times 10^{-3}$ \citep{KochanekEA01} made with the same diet Salpeter IMF that is used in the models. The integral of the predicted star formation rate density over the Hubble time is $\Omega_{\star} = 5.1 \times 10^{-3}$, since a fraction $f_{\rec}=0.33$ of stars terminate in supernovae.

\subsubsection{Stellar mass function}
\label{sec:MFvsz}

A successful galaxy formation model must correctly describe the evolution of the galactic population, while simultaneously matching the local observable properties such as the mass function. This has traditionally proved very difficult, with most models either reproducing the large number of massive galaxies at high redshift but significantly overpredicting the bright end of the local luminosity function (e.g. Kauffmann \etal\/ 1999a,b); or conversely matching the local K-band luminosity function but underpredicting the bright galaxy counts at $z \gtrsim 0.5$ (e.g. Cole \etal\/ 2000; Baugh \etal\/ 2005).

\begin{figure*}[tbph]
\centering
  \subfigure[$z=0.0$]{\includegraphics[height=0.45\textwidth,angle=270]{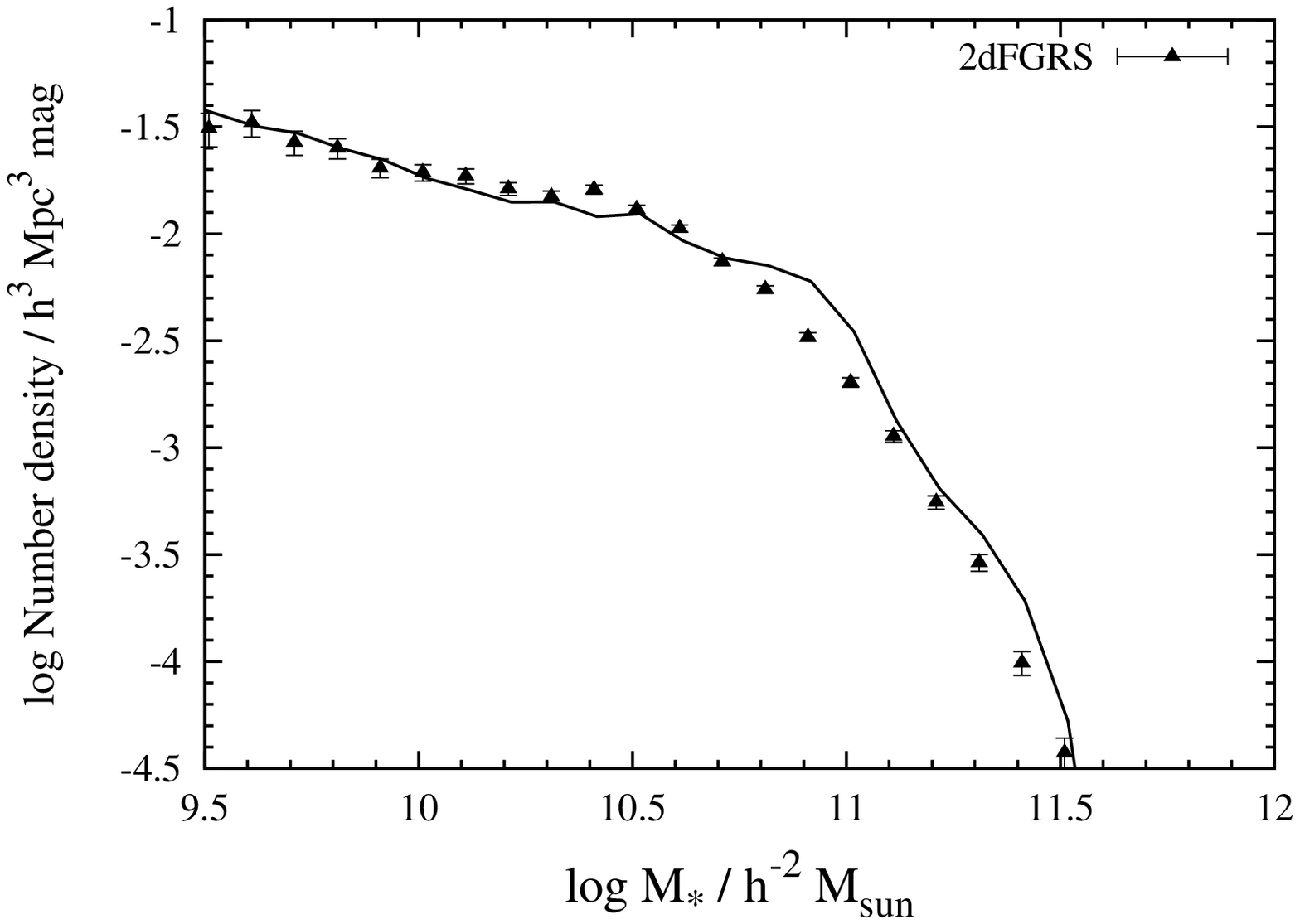}}
  \subfigure[$z=0.5$]{\includegraphics[height=0.45\textwidth,angle=270]{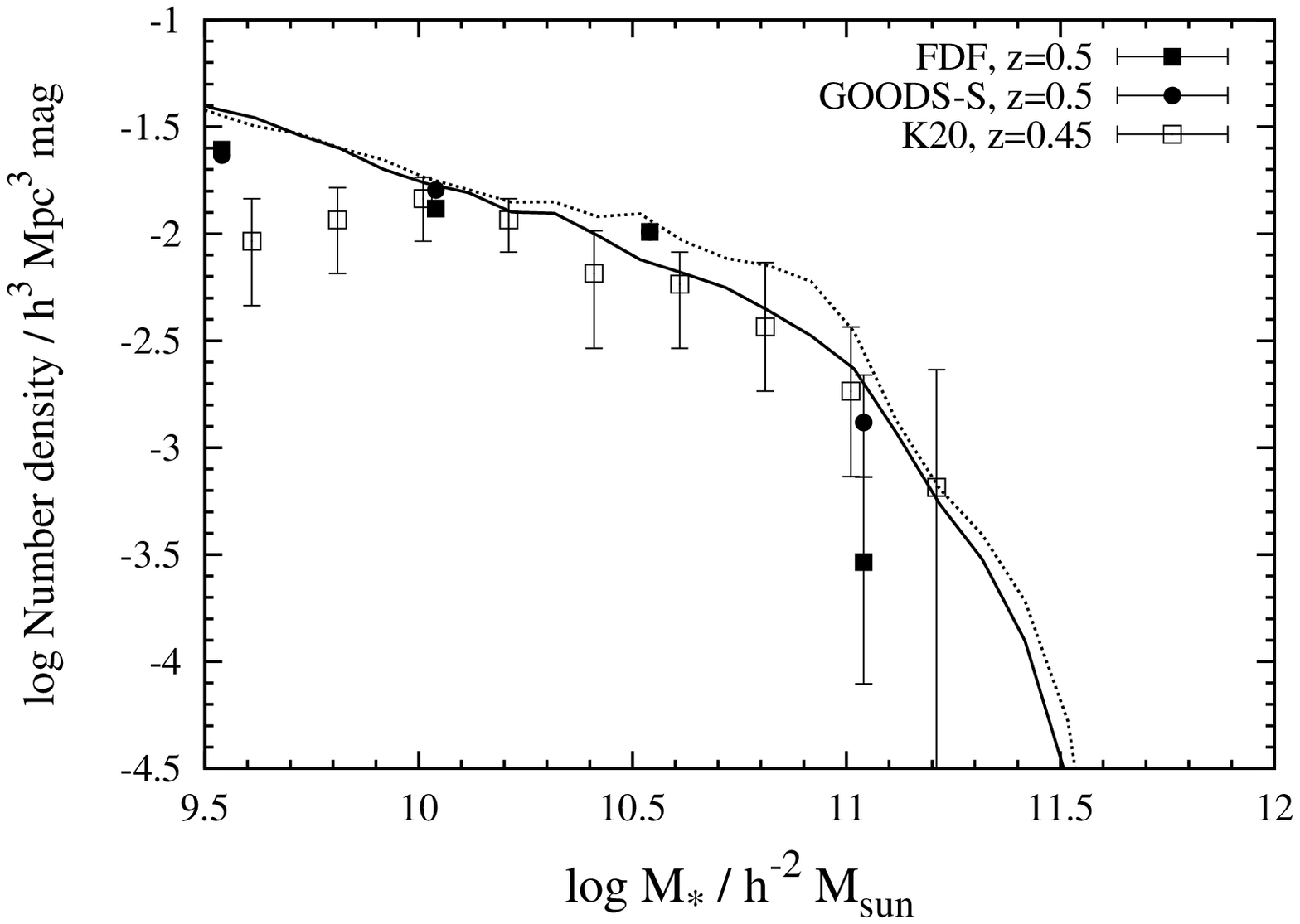}}
  \subfigure[$z=1.0$]{\includegraphics[height=0.45\textwidth,angle=270]{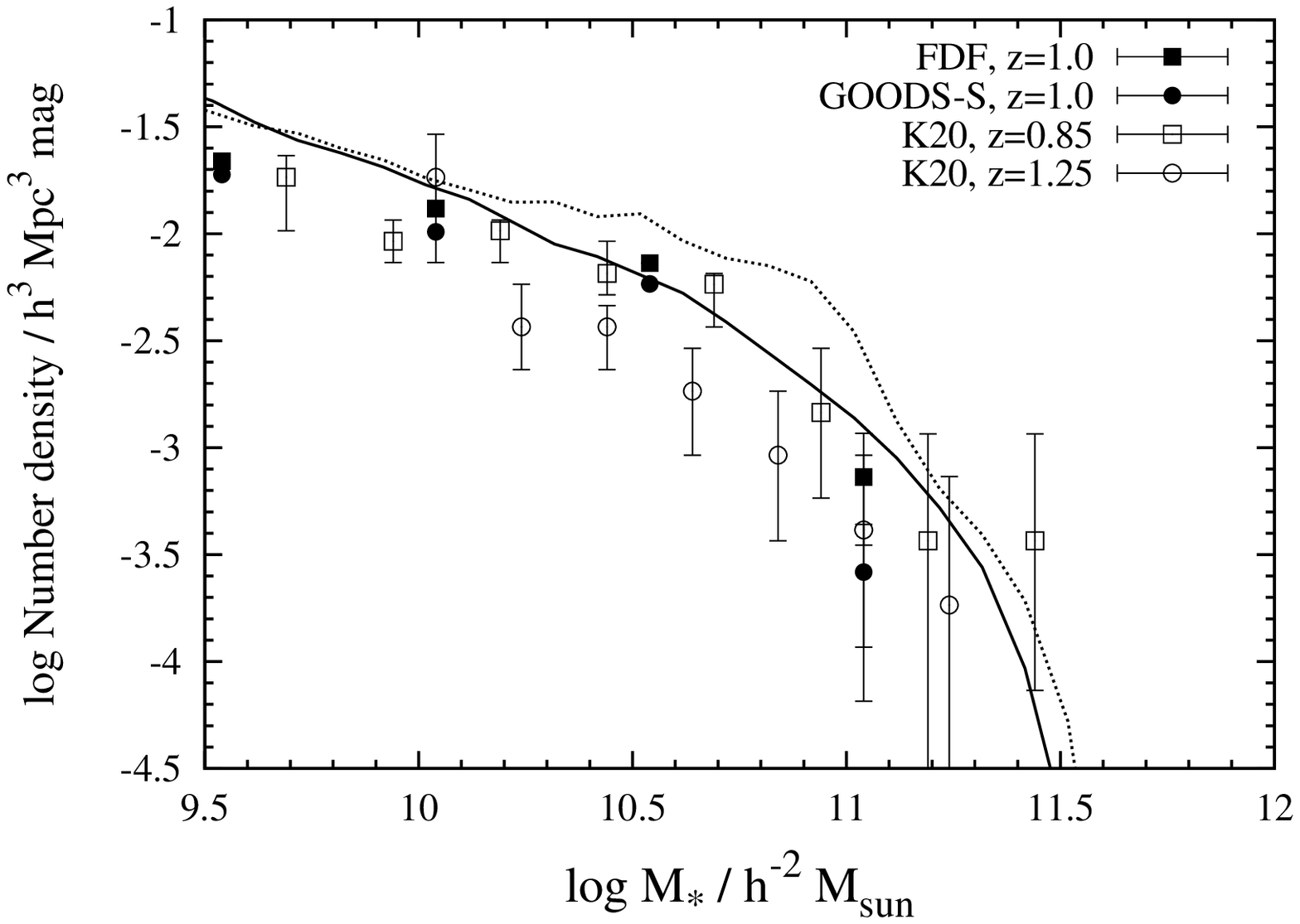}}
  \subfigure[$z=1.5$]{\includegraphics[height=0.45\textwidth,angle=270]{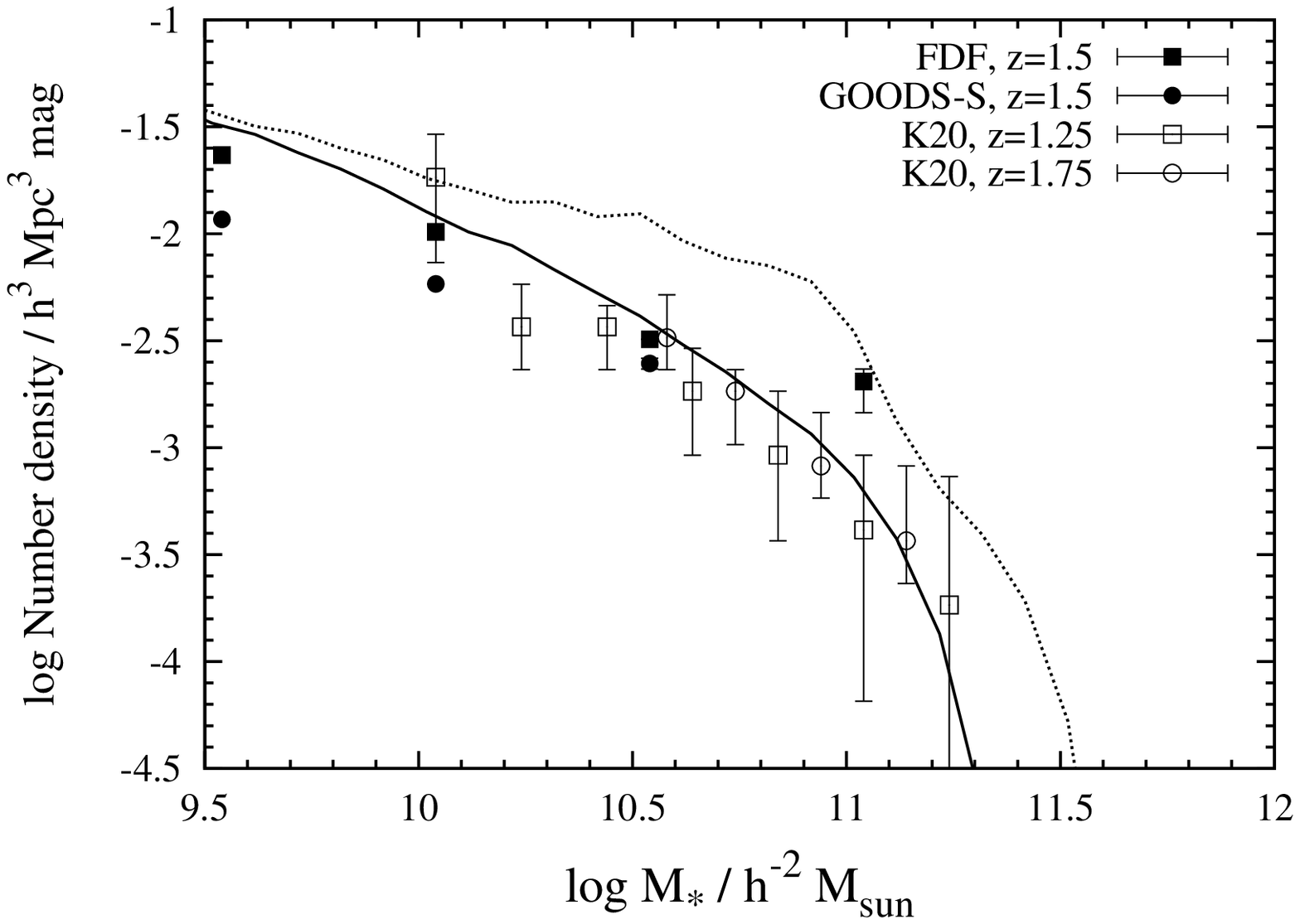}}
  \subfigure[$z=2.0$]{\includegraphics[height=0.45\textwidth,angle=270]{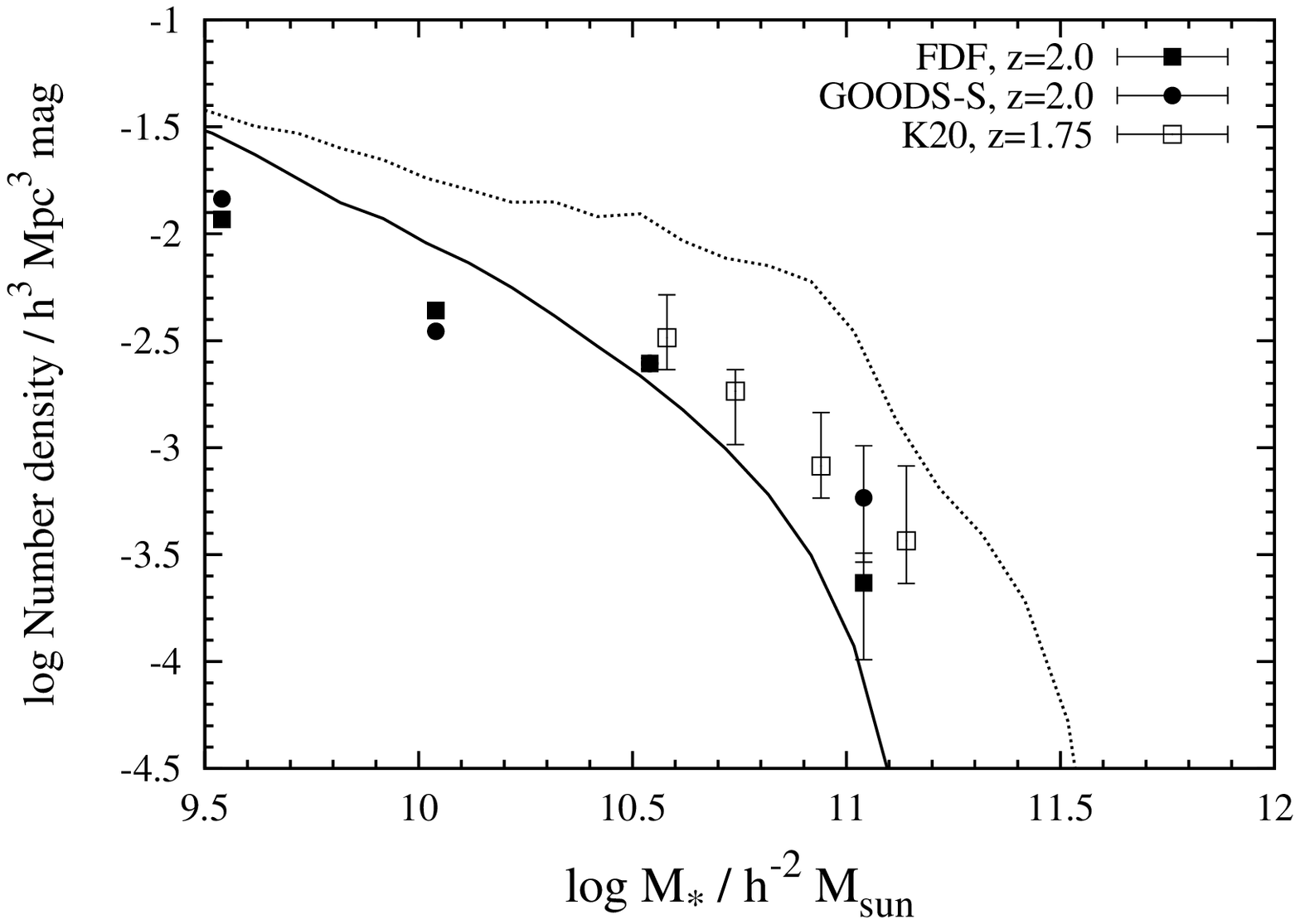}}
  \subfigure[$z=2.6$]{\includegraphics[height=0.45\textwidth,angle=270]{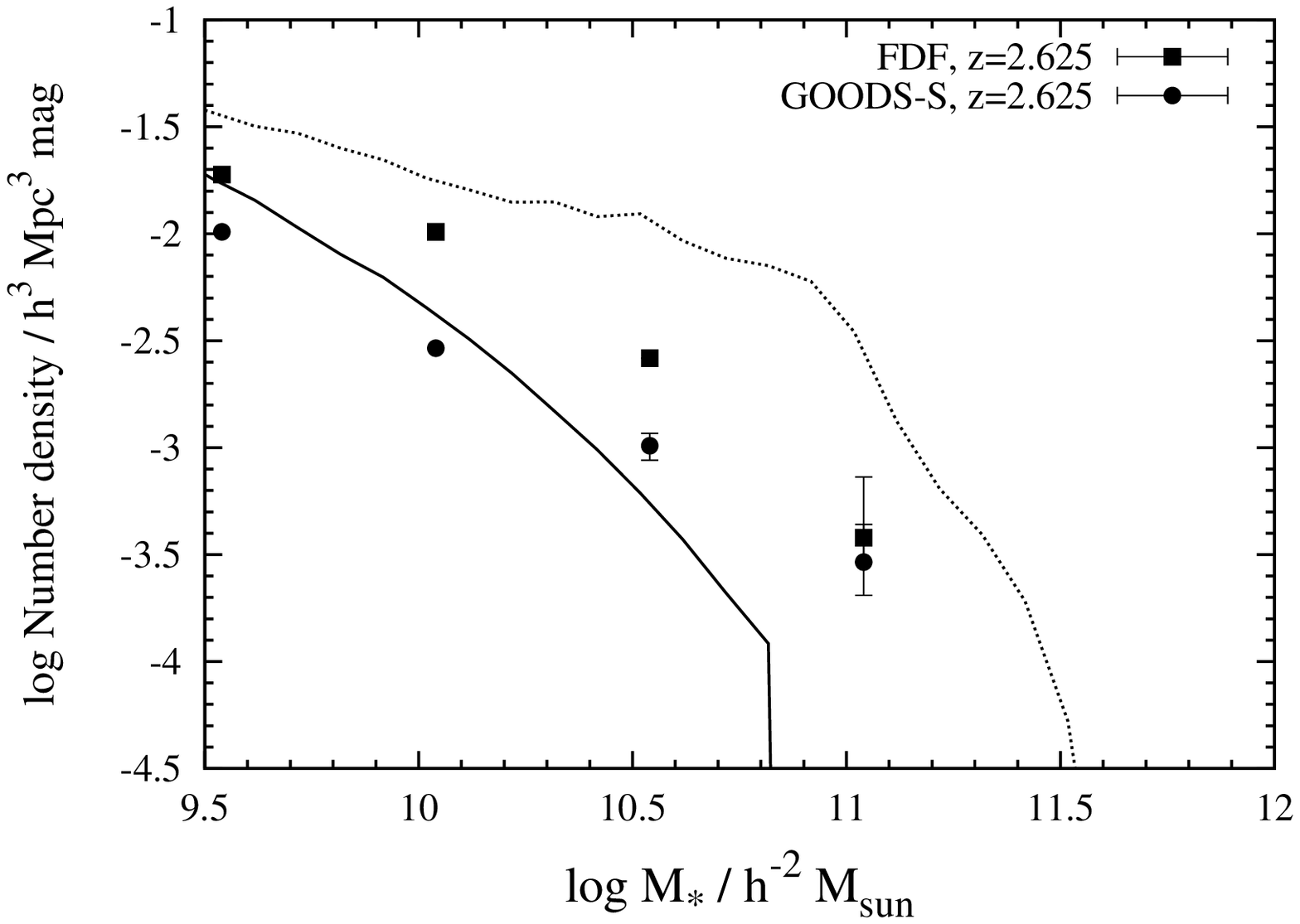}}
\caption{Evolution of the stellar mass function (solid lines). The corresponding quantity at $z=0$ is shown as a dotted line for comparison. Observed data points are derived from multiwavelength photometry and use of Bruzual \& Charlot (2003) spectral synthesis models with a diet Salpeter IMF. Filled triangles are from Cole \etal\/ (2001); filled squares and circles represent Drory \etal\/ (2005) GOODS-South and FORS Deep Field observations; while the K20 survey of Fontana \etal\/ (2004) is shown by open symbols.}
\label{fig:paper_MFvsz}
\end{figure*}

Figure~\ref{fig:paper_MFvsz} plots the evolution of the stellar mass function predicted by the model. It is in remarkable agreement with the results of the K20 survey \citep{FontanaEA04}, and those of the FORS Deep Field (FDF) and GOODS South \citep{DroryEA05} surveys, out to a redshift of $z \gtrsim 1.5$. The only other model in the literature to date providing a similar match is that of Bower \etal\/ (2006).
\begin{figure*}[tbph]
\centering
  \subfigure[$z=1.0$]{\includegraphics[height=0.45\textwidth,angle=270]{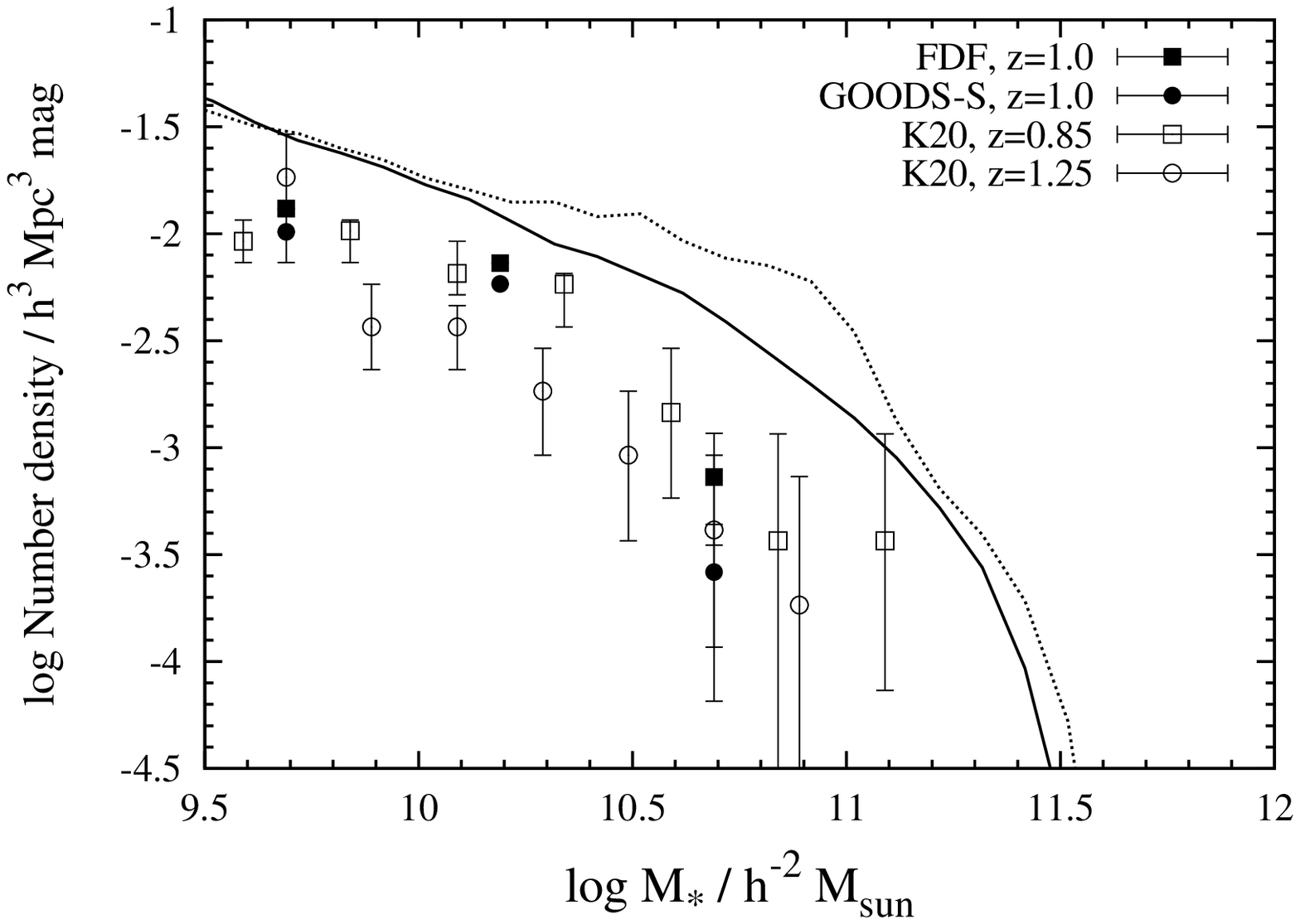}}
  \subfigure[$z=1.5$]{\includegraphics[height=0.45\textwidth,angle=270]{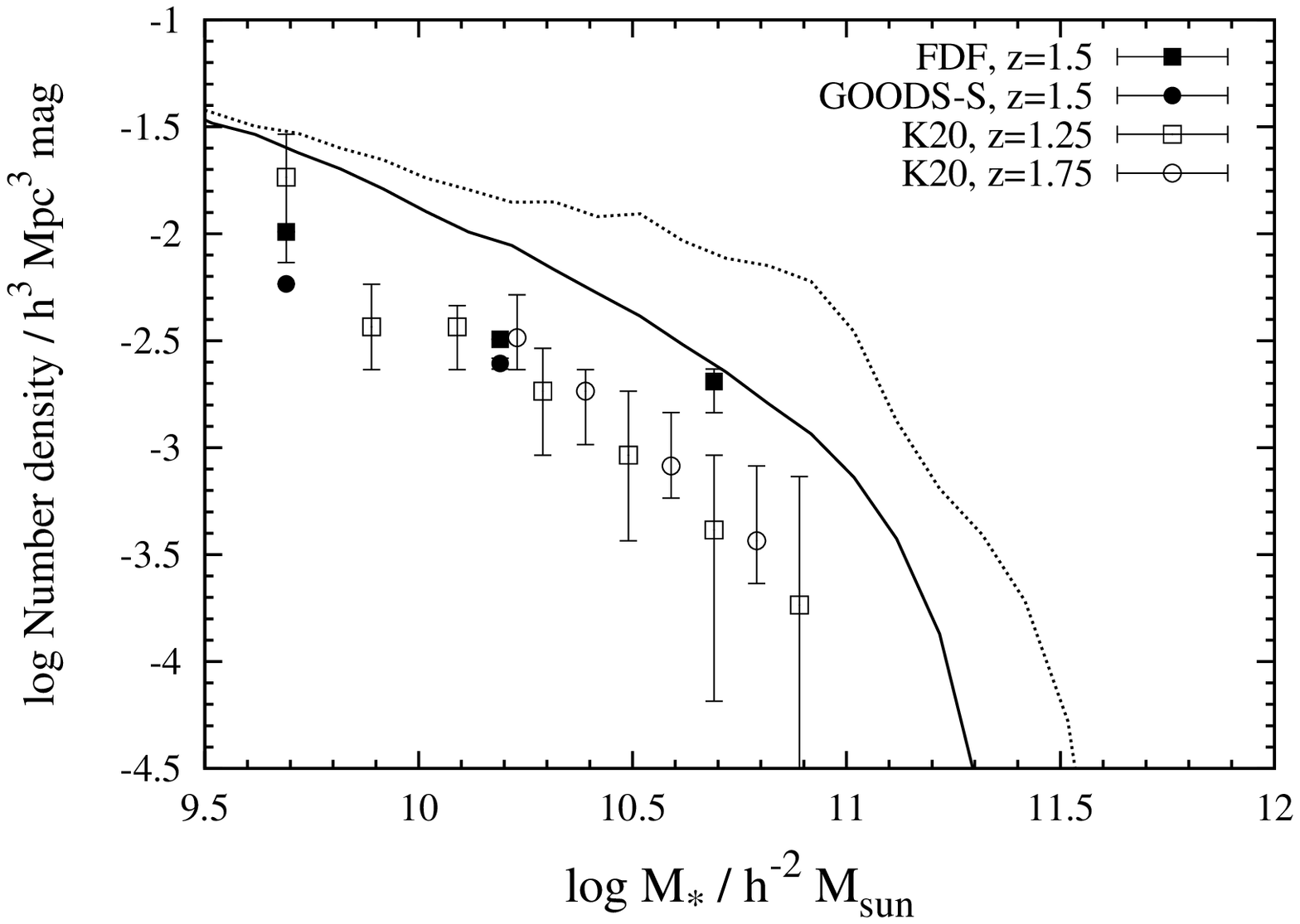}}
  \subfigure[$z=2.0$]{\includegraphics[height=0.45\textwidth,angle=270]{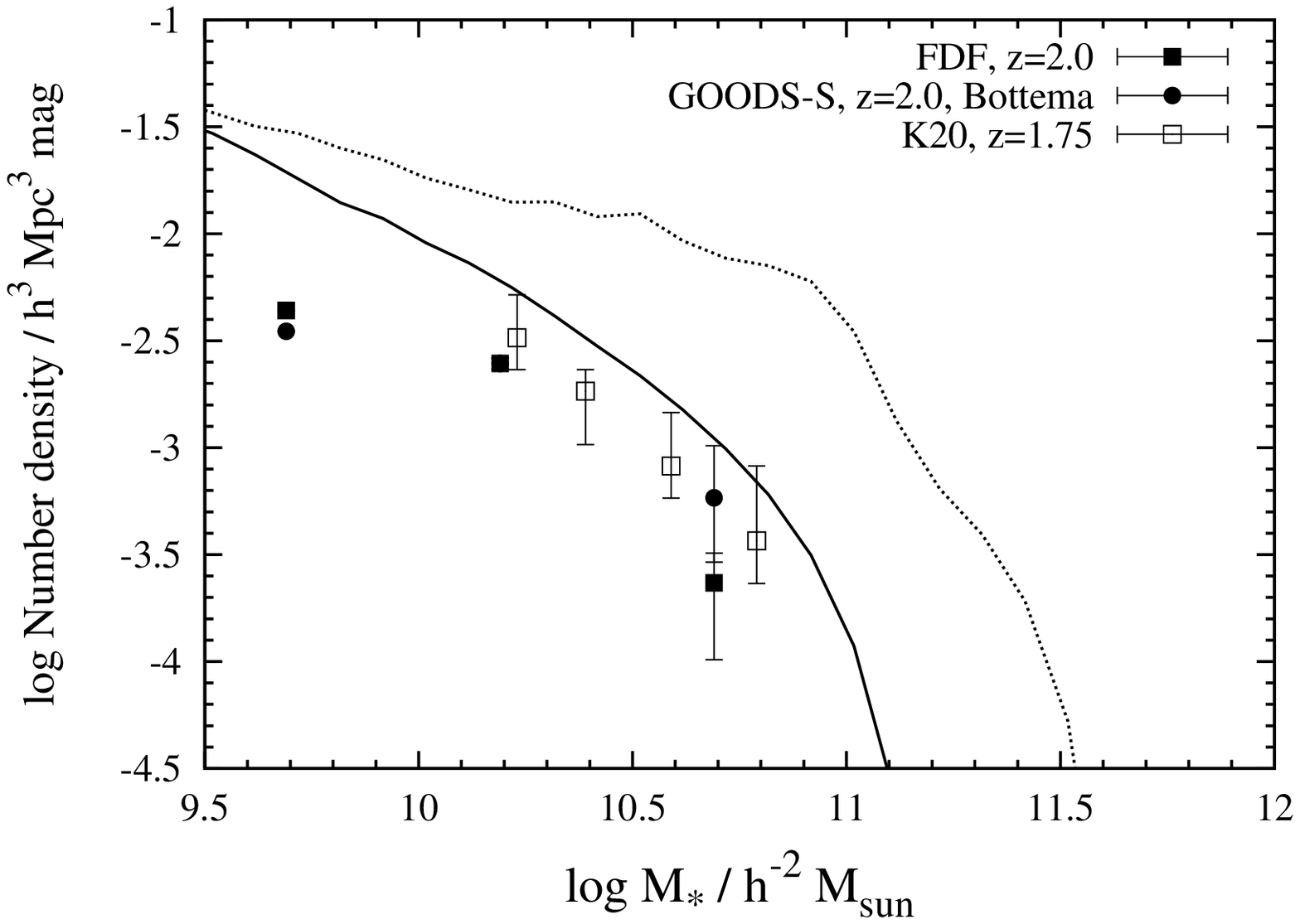}}
  \subfigure[$z=2.6$]{\includegraphics[height=0.45\textwidth,angle=270]{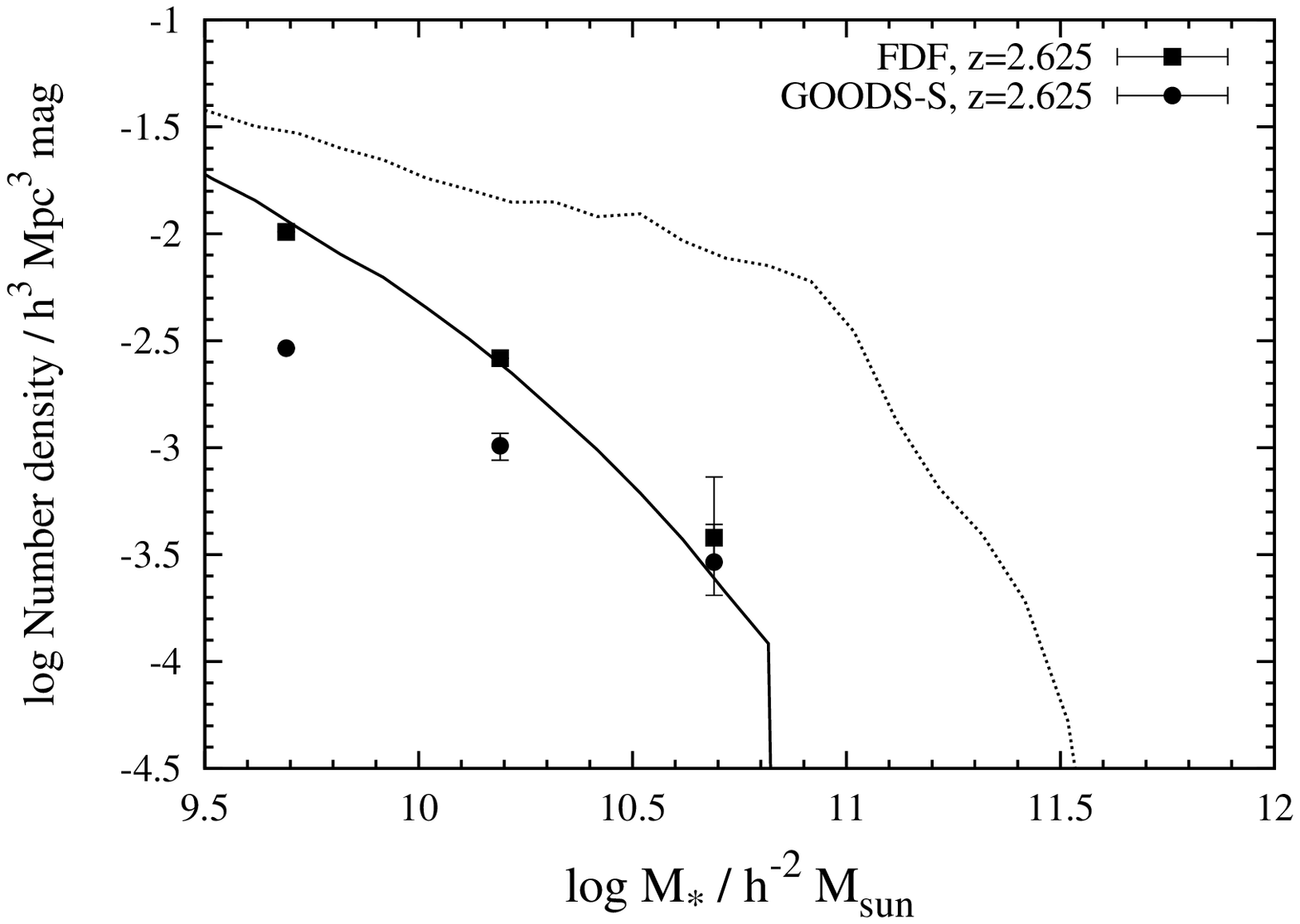}}
\caption{Same as Figure~\ref{fig:paper_MFvsz} but for a Bottema IMF. A change to a top-heavy IMF is required for $z \geq 2$ in order to match observations.}
\label{fig:paper_MFhighzBottema}
\end{figure*}

The last two panels of Figure~\ref{fig:paper_MFvsz} illustrate that the model consistently underpredicts galaxy counts at $z \geq 2$. The uncertainty in the adopted universal IMF cannot account for this as there is a good match between the model and observations at low redshift. However, observations of sub-mm and Lyman break galaxies \citep{BaughEA05} as well as element abundances \citep{NagashimaEA05} suggest that a more top-heavy IMF is required at $z \gtrsim 2$, yielding lower mass-to-light ratios and hence rendering the derived high-redshift stellar masses as overestimates. The physical motivation for two different IMFs comes from a change in the dominant star formation mechanism with cosmic epoch. At high redshifts star formation is mainly triggered by mergers; while at the present epoch it follows the standard disk formation scenario \citep{Kauffmann96,Kennicutt98}.

Figure~\ref{fig:paper_MFhighzBottema} shows the evolution of the stellar mass function under the assumption of a Bottema IMF. This gives good agreement at high redshift, but overpredicts the stellar content at low $z$. These results suggest a switch from the diet Salpeter IMF to a more top-heavy Bottema IMF at $z \gtrsim 2$. Baugh \etal\/ (2005) estimate that quiescent star formation (i.e. in disks) dominates for $z \lesssim 2-3$, consistent with our findings.

The slight overprediction in the counts of low-mass galaxies at $z>0$ is consistent with the results of other models (e.g. Bower \etal\/ 2006). There are two possible causes of this. One possibility would be current implementations of supernovae feedback truncating star formation too early. Alternatively, observational constraints at the low-mass end are likely to suffer from incompleteness, as evidenced by substantial field-to-field variations in Figures~\ref{fig:paper_MFvsz} and \ref{fig:paper_MFhighzBottema}.

\subsection{Cosmic downsizing}
\label{sec:cosmicDownsizing}

From Figures~\ref{fig:paper_MFvsz} and \ref{fig:paper_MFhighzBottema} it is clear that the high-mass end of the local stellar mass function is assembled at $z>1$, while lower-mass objects continue forming stars until the present day. This is consistent with observations of the most massive galaxies having red colours \citep{PanterEA07} and in stark contrast to the hierarchical assembly of halo mass.

\begin{figure*}[tbph]
\centering
  \subfigure[AGN feedback]{\includegraphics[height=0.45\textwidth,angle=-90]{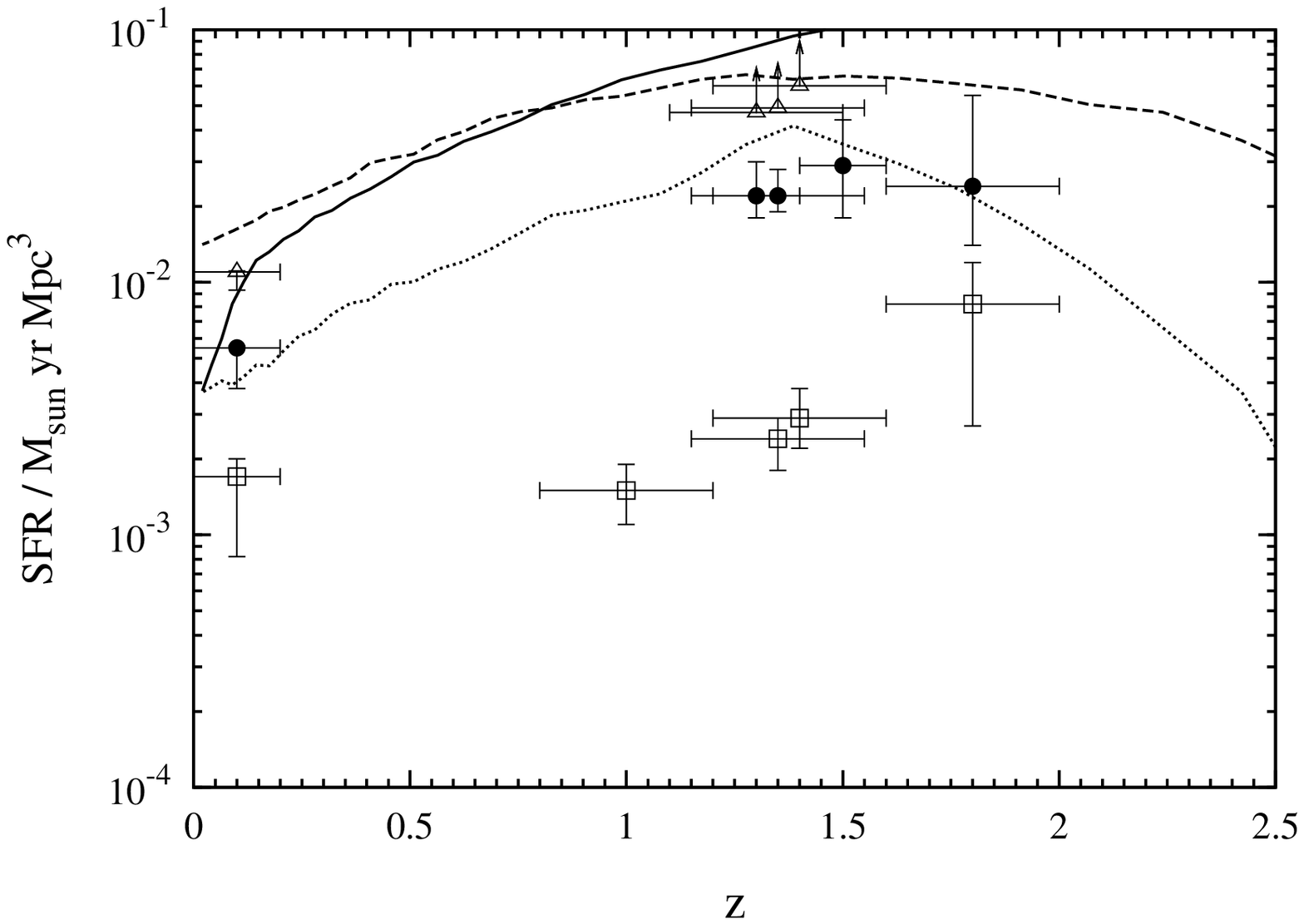}}
  \subfigure[No AGN]{\includegraphics[height=0.45\textwidth,angle=-90]{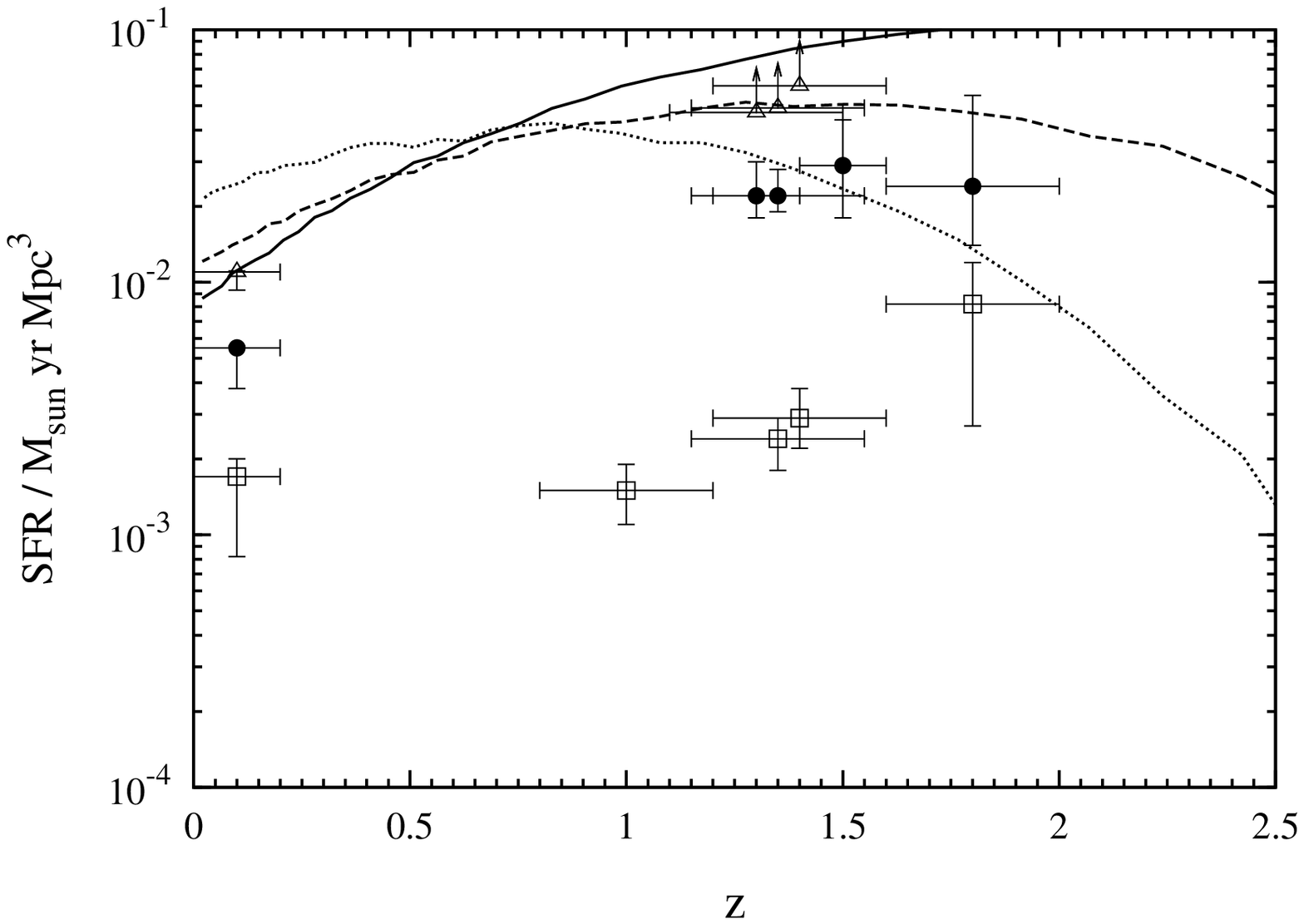}}
\caption{Star formation rate densities broken up by stellar mass for the best-fit (left) and no-AGN (right panel) models. All curves are for: $10^{9.26}<\frac{M_{\star}}{\rm M_\odot}<10^{10.46}$ (solid line, open triangles), $10^{10.46}<\frac{M_{\star}}{\rm M_\odot}<10^{11.06}$ (dashed line, closed circles), $10^{11.06}<\frac{M_{\star}}{\rm M_\odot}<10^{11.76}$ (dotted line, open squares). Data points for $z>0.5$ are from Juneau \etal\/ (2005); and from Brinchmann \etal\/ (2004) for the local volume. A colour version of this Figure is available in the online article.}
\label{fig:paper_SFRvsz_byMstars}
\end{figure*}

The left panel of Figure~\ref{fig:paper_SFRvsz_byMstars} shows the evolution of star formation rate density as a function of stellar mass. The stellar mass bins are chosen to follow the observations of Juneau \etal\/ (2005) and Brinchmann \etal\/ (2004) converted to a diet Salpeter IMF. The same quantities for a model with no AGN feedback are plotted in the right panel.

Figure~\ref{fig:paper_SFRvsz_byMstars} shows that inclusion of AGN feedback changes the galaxy assembly paradigm from hierarchical to anti-hierarchical. There are three related reasons for the preferential suppression of star formation in massive galaxies at low redshift. Firstly, lower densities in the low-$z$ Universe mean the ratio of cooling time to free-fall time is higher there than at high $z$, making AGN feedback more effective at low redshifts. Furthermore, black holes are more massive at low $z$, allowing higher jet powers to be generated by ADAF disks (Equation~\ref{eqn:QjetADAF}). Apart from providing greater shock heating, this also introduces an additional mechanism of star formation suppression via gas ejection by powerful radio sources \citep{BassonAlexander03}. The absence of this mode of feedback in the presented model is most likely responsible for the disagreement between model predictions and observations in intermediate and massive galaxies and at low $z$.

\begin{figure*}[tbph]
\centering
  \subfigure[mass-weighted]{\includegraphics[height=0.45\textwidth,angle=-90]{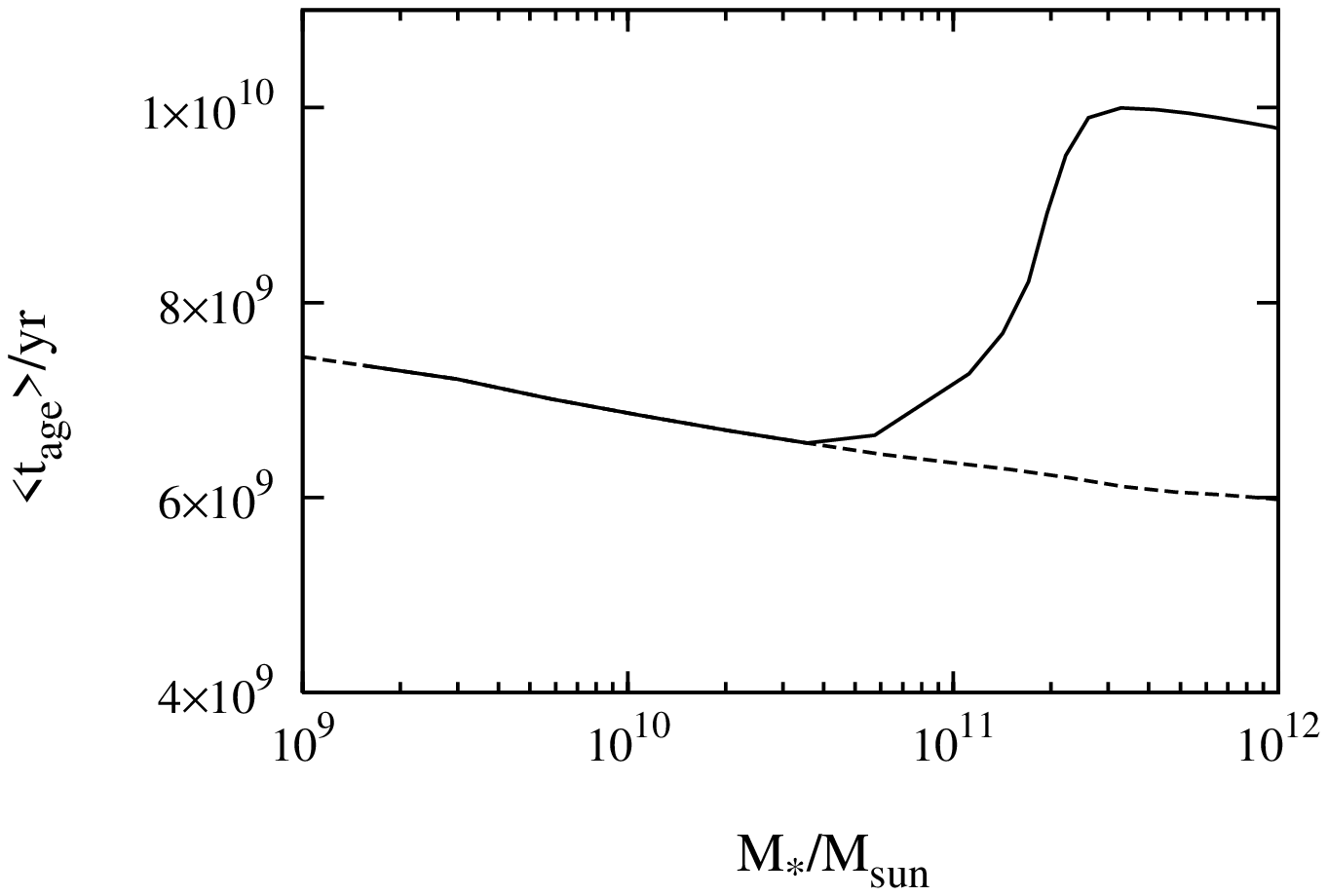}}
  \subfigure[luminosity-weighted]{\includegraphics[height=0.45\textwidth,angle=-90]{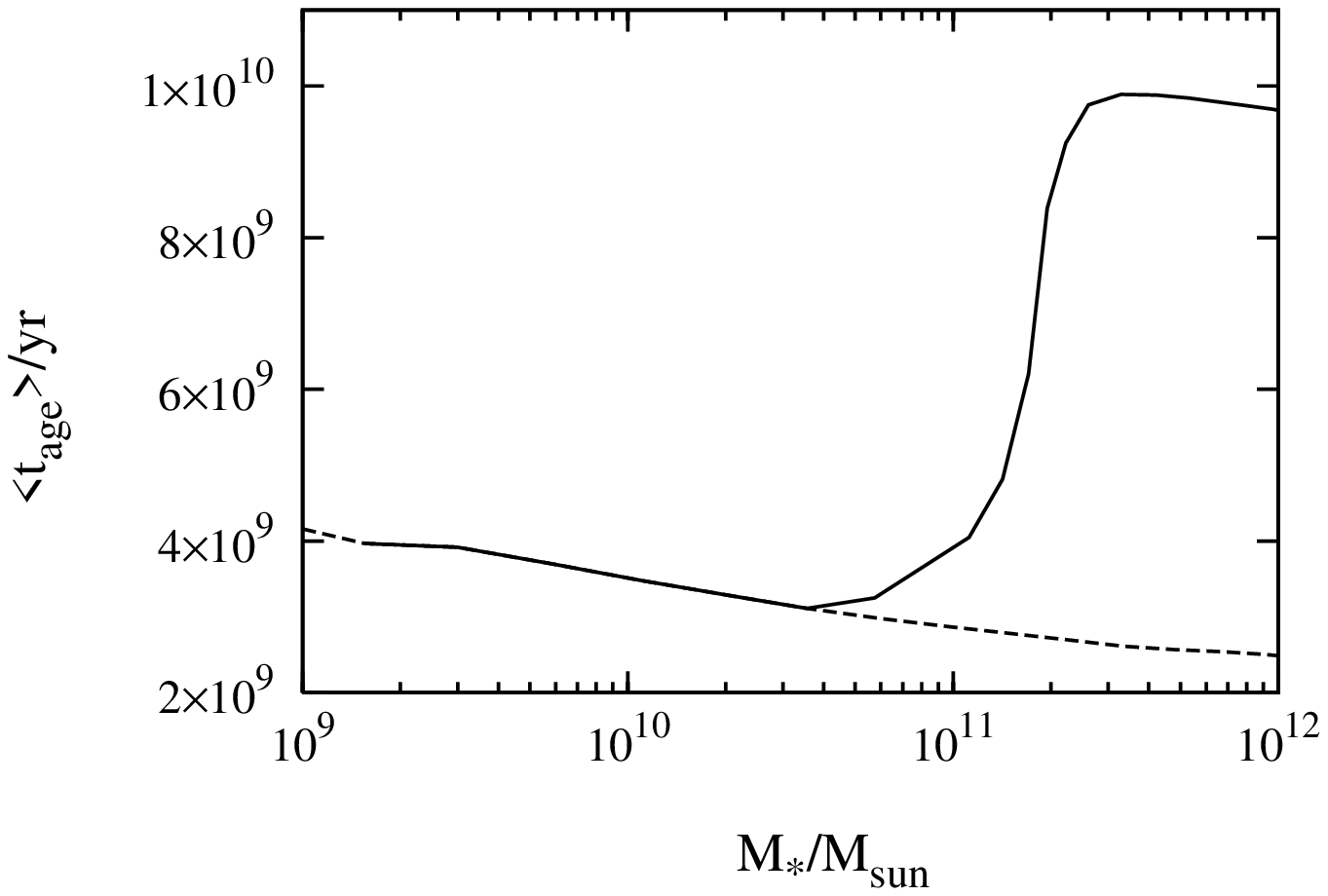}}
\caption{Mean stellar ages weighted by mass (left) and luminosity (right panel) for models with reionization and supernovae feedback but no AGN (dashed); and the full model including radio source heating (solid curve). When AGN feedback is included the population is bimodal, with the most massive galaxies containing oldest stars.}
\label{fig:paper_meanStellarAge}
\end{figure*}

A final illustration of radio source heating giving rise to a massive, red population of galaxies is given in Figure~\ref{fig:paper_meanStellarAge}. Here the mean mass- and luminosity-weighted stellar ages at the present epoch are plotted as a function of galaxy stellar mass. Results for a model with no AGN, but including gas cooling, and reionization and supernovae feedback are also shown for comparison. In the absence of any feedback the most massive galaxies host the youngest stars, and hence are blue, as expected from the standard hierarchical formation models. The situation changes dramatically when radio source heating is introduced, with galaxies more massive than $10^{11}$~\Msun\/ at $z=0$ containing older stars than predicted by the no-AGN case. Such bimodality is consistent with the semianalytical model results of Croton \etal\/ (2006) and Bower \etal\/ (2006) and, significantly, the observed sharp break between old and young stellar populations at $M_{\star} \sim 3 \times 10^{10}$~\Msun\/ in SDSS galaxies \citep{KauffmannEA03b}.

\section{Summary}

In this work, we present a new galaxy formation and evolution model. Gas is assumed to be accreted onto the halo along with the dark matter, and distributed in a quasi-isothermal fashion. Gas cooling competes with feedback heating from the ionizing UV background, supernovae and radio sources. Low-mass haloes dominate the stellar mass budget of the Universe at the present epoch, and efficiency of supernovae feedback is tightly constrained by the star formation history of the Universe and counts of low-to-intermediate-mass galaxies. The observed local correlation between central black hole masses and those of stars in the bulge sets the rate at which gas is accreted by the black hole.

For the first time, we implement realistic intermittent AGN feedback resulting from shock heating by expanding radio sources. Radio source heating is found to suppress star formation in the most massive galaxies, providing an excellent match to the observed local stellar mass function. Observations of the stellar mass function at earlier epochs suggest the stellar IMF may change qualitatively around $z=2$ from a merger-dominated, top-heavy IMF to one consistent with observations of the local Universe.

The model reproduces the anti-hierarchical paradigm of structure formation, with the top end of the stellar mass function in place by $z \sim 1.5$. Also predicted is the sharp observed transition in the mean stellar ages of galaxies at $M_{\star} \sim 2-3 \times 10^{10}$~\Msun\/, with the more massive structures being ``red and dead''. The self-regulatory nature of this AGN feedback channel is illustrated by our model reproducing the local black hole - spheroid mass relation, and suggests that intermittent radio source feedback can provide the missing link in galaxy formation and evolution models.

\section{Acknowledgements}

We thank Philip Best for useful discussions, and the anonymous referee for a number of suggestions that have helped improve the paper. SS is grateful to the Cambridge Commonwealth Trust and the Isaac Newton Trust for support. This work has made use of the Distributed Computation Grid of the University of Cambridge (CamGRID).

\end{document}